\documentclass[11pt,reprint,aps,prb,amsmath,amssymb,longbibliography,floatfix,nobibnotes]{revtex4-2}
\usepackage{natbib}
\usepackage{graphicx}% Include figure files
\usepackage{dcolumn}% Align table columns on decimal point
\usepackage{bm}% bold math
\usepackage{fix-cm} % Allow scalable Computer Modern fonts
\usepackage{hyperref}
\hypersetup{
    colorlinks=true,
    linkcolor=black,
    citecolor=black,
    urlcolor=black,
    pdfpagemode=FullScreen,
}

\usepackage{pgfplots}
\usepackage{tikz}
\usepackage{amsmath}
\usepackage{amssymb}
\usepackage{comment}
\usepackage{siunitx}
\usepackage{physics}
\usepackage{xfrac}
\usepackage{relsize}
\usepgfplotslibrary{polar}
\pgfplotsset{compat=1.18,width=6cm}

\newcommand{\cl}{\mathcal{C}}
\newcommand{\cres}{c_{\mbox{res}}\z}
\newcommand{\cresp}{c_{\mbox{res}}^P}
\newcommand{\cresu}{c_{\mbox{res}}^U}
\newcommand{\dg}{^\dagger}
\newcommand{\dl}{\Delta_L\z}

\newcommand{\dn}{\mathcal{D}_{\mathcal{N}}}

\newcommand{\eb}{\epsilon_B\z}

\newcommand{\ff}{\mathcal{F}}
\newcommand{\fp}{{\mathcal{P}}_{P}\z}
\newcommand{\fu}{{\mathcal{P}}_{U}\z}
\newcommand{\ffp}{{\mathcal{F}}_P\z}
\newcommand{\ffu}{{\mathcal{F}}_U\z}
\newcommand{\fr}{{\ff}_{\mbox{ref}}\z}
\newcommand{\fz}{{\ff}_{\zeta, \theta}\z}
\newcommand{\hc}{\mathrm{H.~c.}}
\newcommand{\hh}{\mathcal{H}}

\newcommand{\omb}{\omega_B^{\ }}
\newcommand{\omtau}{\omega_{\tau}^{\ }}

\newcommand{\oml}{\omega_L\z}
\newcommand{\pp}{\mathcal{P}}
\newcommand{\rt}{\mathcal{R}_t\z}

\newcommand{\nn}{\mathcal{N}}
\newcommand{\erg}[1]{E_{#1}}
\newcommand{\rp}{r_P\z}
\newcommand{\ru}{r_U\z}

\newcommand{\seth}{\{h\}}
\newcommand{\setp}{\{p\}}
\newcommand{\setk}{\{k\}}
\newcommand{\sz}{{2L}}
\newcommand{\tf}{\mathcal{T}_t}

\newcommand{\tz}{{T}_\zeta\z}

\newcommand{\z}{^{\phantom{\dagger}}}
\newcommand{\Gusfrac}[2]{\mathlarger{\frac{#1}{#2}}}
\newcommand{\tmax}{t_{\textrm{max}\z}}
\newcommand{\lno}[1]{{#1}}
\newcommand{\fdp}[1]{{#1}}

\DeclareMathOperator{\sii}{Si}
\DeclareMathOperator{\ci}{Ci}

\begin{document}
%\bibliographystyle{apsrev4-2}

%\title{Time-dependent approach to the X-ray photoemission problem}
\title{Temporal structures of the X-ray photoemission problem}
\author{F. D. Picoli¹, G. Diniz$^{1,2}$, M. P. Lenzarini¹, I. D'Amico³, L. N. Oliveira¹}
\altaffiliation[luizno@usp.br]{}%Lines break automatically or can be forced with \\
\affiliation{¹Instituto de Física de São Carlos, Universidade de São Paulo, São Carlos, São Paulo, Brazil \\
²Institute of Physics of the Czech Academy of Sciences, Prague, Czech Republic \\
³School of Physics, Engineering and Technology, University of York, York, United Kingdom,
³York Center for Quantum Technologies, University of York, York, YO10 5DD, United Kingdom
}
\begin{abstract}
{Theoretical studies of X-ray photoemission from simple metals have traditionally focused on the frequency domain, aiming to reproduce experimental spectra.  Here, we investigate the same problem in the time domain in search of physical insight and methodological advances. Our results reveal prominent aspects of the problem that are inconspicuous in the frequency domain. The calculated $\mathcal{F}(t)$
exhibits a weakly damped harmonic oscillation that modulates the Doniach-Sunjic power-law decay of the photoemission rate, a behavior arising from the coherent interference between two classes of particle-hole excitations. From a methodological perspective, we advance the time-dependent numerical renormalization-group (NRG) approach by exploring the eNRG method, a real-space variant that is more flexible than Wilson's construction. Giving special attention to strong core-hole potentials, we compare the photocurrents obtained from two complementary time-dependent eNRG algorithms with (i) an analytical expression for $\mathcal{F}(t)$ that becomes highly accurate at moderately long times and (ii) results from numerical diagonalization of a tight-binding Hamiltonian, which covers the time interval in which our analytical expression is less precise.
Anticipating extensions to correlated-impurity models, we identify the sources of deviation and discuss the virtues and drawbacks of the two algorithms.}
\end{abstract}
\maketitle
\newpage

\section{\label{sec:level1}Introduction}

In this age of algorithms, methods specially designed to deal with time dependence have enriched condensed matter theory. Experimental progress toward understanding nonequilibrium properties \cite{Cui2014,PhysRevB.59.10935,Ferrier2016,CHERNEIII2001985}, ultrafast phenomena \cite{PhysRevLett.109.087401,PhysRevLett.111.153602,Foster2008,2010SFKp103003,2018Mil8}, mesoscopic dynamics \cite{doi:10.1080/00107518508210991,TJThornton_1995,Reguera2005}, and other enigmatic features of the time domain has aroused interest in such methods \cite{Fujisawa_2006,PhysRevB.85.085113,https://doi.org/10.1002/cphc.201301137}. The most successful numerical procedures are offshoots of time-honored techniques that have proven successful in the computation of equilibrium properties and share their virtues and limitations with the parent methods. Time-Dependent Density Functional Theory (TD-DFT) \cite{PhysRev.140.A1133,PhysRevLett.52.997,Marques2004,Marques2006,Ullrich2012} and Time-Dependent Density Matrix Renormalization Group (tDMRG) \cite{PhysRevLett.69.2863,PhysRevLett.93.076401,PhysRevB.78.195317} offer good illustrations. TD-DFT is remarkably successful; for instance, it has extended the domain of DFT beyond just the ground-state densities. Similarly to DFT, the time-dependent version relies on the quality of the available density functionals. While it can handle large systems in one, two, and three dimensions, its application is primarily limited to weakly correlated systems.

%tDMRG emerged from DMRG to provide insight into the dynamical and nonequilibrium properties of strongly correlated systems \cite{PhysRevLett.93.076401,Trotzky2012,Hofstetter_2018,Zawadzki2023}. Parallel to the breadth of the parent method, the reach of tDMRG is limited by the dimensionality and size of the object under study \cite{https://doi.org/10.1002/ctpp.201610003,grundner2024complex}. This limitation is practically irrelevant for the typical unidimensional Hamiltonian since sufficiently long lattices fit even in small computational budgets.

%\fdp{tDMRG emerged from DMRG to provide insight into the dynamical and nonequilibrium properties of strongly correlated systems \cite{PhysRevLett.93.076401,Trotzky2012,Hofstetter_2018,Zawadzki2023}. Parallel to the breadth of the parent method, the reach of tDMRG is limited by the growth of entanglement during time evolution, referred to as the entanglement barrier, which restricts the accessible simulation times and effective system sizes. This limitation becomes more severe with increasing dimensionality and system size \cite{Schollwock2011,https://doi.org/10.1002/ctpp.201610003,grundner2024complex}.}

tDMRG emerged from DMRG to provide insight into nonequilibrium properties of strongly correlated systems \cite{PhysRevLett.93.076401,Trotzky2012,Hofstetter_2018,Zawadzki2023}. Parallel to the breadth of the parent method, {the reach of tDMRG is limited by the growth of entanglement, which can restricts the accessible simulation times. This limitation becomes more severe with increasing dimensionality and system size \cite{Schollwock2011,https://doi.org/10.1002/ctpp.201610003,grundner2024complex}.}

%There are exceptions to this.

{In particular}, low-energy particle-hole excitations play a crucial role in determining the physical properties of specific models, including those that exhibit Kondo physics \cite{Wilson_NRG,1998GSM+156,PhysRevLett.108.086405,PhysRevB.101.125115,OBrien2014,2020BSCp210,2021XSKp087101,PhysRevB.106.125413}. 
%In such cases, computing equilibrium properties can be challenging. 
The very long lattices required to achieve the necessary energy resolution often require extensive computational runs, which can be prohibitively expensive {for DMRG}. The numerical renormalization group (NRG) method \cite{Wilson_NRG,1980KWW1003,Bulla,bulla2001finite} offers a promising alternative, as it can handle very long lattices at a relatively low computational cost. Computations of ground-state, thermodynamic, transport, and frequency-dependent excitation properties of models for correlated impurities coupled to noninteracting electrons have thoroughly tested the NRG method \cite{PhysRevB.75.041307,PhysRevB.90.045117,PhysRevB.81.115316,PhysRevLett.47.1553,OLIVEIRA199161,PhysRevB.81.235127,PhysRevB.66.174404}.

Its record in the time domain is less extensive. Early attempts to compute time dependencies faced unexpected difficulties. Generalizations of Wilson's procedure were brought forward later, with a view to overcoming such obstacles. In this context, Anders and Schiller \cite{2005AnS196801,2006AnS245113} pointed out that high-energy excitations affect dynamics even at long times and showed a way to overcome this complication. Costi and co-workers \cite{nghiem2014generalization,nghiem2014time} developed a special procedure to deal with pulses. More recent developments have proposed two hybrid constructions. One of them combines logarithmic and linear discretizations to study nonequilibrium transport \cite{Schwarz2018,lotem2020}. The second generalizes the NRG method by introducing thermal reservoirs to compensate for the high-energy degrees of freedom excluded from the logarithmic discretization procedure \cite{Bruognolo2017}.

Here, moved by the desire to preserve the simplicity of the original procedure, we apply a time-dependent real-space version of Wilson's approach to the model that supported the development of the frequency-dependent procedure \cite{1981OlW4863}. Specifically, our testbed is photoemission from a simple metal, and we examine the time dependence of the photoemission current. Preliminarily, we combine analytical and numerical treatments to calculate the time dependence for a tight-binding model. We thoroughly discuss the physics of the results and save them as a reference. We then present two renormalization-group procedures based on the eNRG procedure, the real-space formulation proposed in~\cite{2022FeO075129}, and compare the resulting photoemission currents.

Our report comprises six sections. It starts with an overview of the findings in Sec.~\ref{sec:sum_findings}. Next, Sec.~\ref{sec:model} defines the model based on a tight-binding Hamiltonian. Section~\ref{sec:fidel} starts with a calculation to set benchmarks to gauge the precision of the renormalization-group results. In Sec.~\ref{sec:numerical-results}, a straightforward numerical procedure, combined with analytical treatment, accurately determines the time-dependent photocurrent for the tight-binding model. We interpret the results physically. Section~\ref{sec:enrg} describes the eNRG method, presents two alternative procedures to calculate the photocurrent, and discusses the deviations from the reference curves obtained in Sec.~\ref{sec:fidel} as functions of the eNRG parameters. Finally, Sec.~\ref{sec:conclude} gathers our conclusions.

\section{Summary of findings}\label{sec:sum_findings}
Photoemission is a relatively simple phenomenon. X radiation ejects a core electron, leaving behind a hole whose potential attracts the conduction electrons. {Experimentalists} measure the photoemission current as a function of energy. The resulting spectrum shows a peak at the threshold energy, the minimal energy required to photoeject the core electron. The peak is asymmetric and somewhat broad. The asymmetry reflects the creation of electron-hole pairs above the threshold, while the breadth reflects the finite lifetime of the core hole.

In contrast, we focus on the time dependence of the photocurrent, given by the squared projection of the quantum state $\ket{\Psi(t)}$ on the ground state before the x-ray creates the core hole. Insofar as this ground state constitutes our initial state $\ket{\Psi(0)}$, the photocurrent is mathematically equivalent to the \emph{fidelity} $\ff(t)$ of $\ket{\Psi(t)}$ relative to its value at $t=0$. X-ray photoemission spectroscopy (XPS)  measures the Fourier transform of the fidelity \cite{GRECZYNSKI2020100591,PhysRevLett109087401,Mahan2010-xj}. {X-ray absorption spectroscopy (XAS) complements XPS by measuring the probability of promoting a core electron to the conduction band \cite{Mahan2010-xj,Chantler2024}}.

Interest in XAS and XPS motivated mathematical studies in the late 1960s \cite{mahan1967excitons,1969NoD1097,1970DoS285}, which identified power laws with universal exponents in the near-threshold frequency dependence of the currents. The core potential shifts the phases of the conduction electrons, and the phase shift $\delta$ of the conduction state at the Fermi level parameterizes the power laws. Nozières and De Dominicis determined the XAS exponent \cite{1969NoD1097}
\begin{align}
  \label{eq:53}
  \beta \equiv 2\left(\dfrac{\delta}{\pi}-1\right)^2,
\end{align}
and Doniach and Sunjic found the XPS exponent \cite{1970DoS285}
\begin{align}
  \label{eq:39}
  \alpha = 2\left(\dfrac{\delta}{\pi}\right)^2.
\end{align}

Subsequent research determined the prefactors of the power laws and identified a secondary threshold at higher energy, associated with excitations from the \emph{bound state} that splits off the bottom of the conduction band under the attraction of the core hole \cite{Nozieres71,ohtaka1990,cornaglia2007}. We are unaware of parallel developments in the time domain.

%As mentioned, XAS and XPS proved instrumental in extending the NRG method to excitation properties in the frequency domain \cite{PhysRevB224994,oliveira1981,cox1985,libero1990,oliveira1985,PhysRevB.24.4863}.  

{XAS and XPS proved instrumental in extending NRG to excitation properties in the frequency domain \cite{PhysRevB224994,oliveira1981,cox1985,libero1990,oliveira1985,PhysRevB.24.4863}}.
In the time domain, the photoemission problem is more than a checkpoint in the algorithm-development itinerary, {exhibiting features that merit physical discussion}. As expected, the fidelity decreases following a Doniach-Sunjic power law, but we shall see that a sharply defined damped oscillation envelops the decay. This behavior evidences interference among transitions from the initial many-body state (the quantum state of the electron gas before absorption of the x-ray) to the final many-body states (the excited states generated by the sudden creation of a localized potential). The occupation of the bound state sorts the final states into two classes: the \emph{plugged} many-body states, with an occupied bound state, and the \emph{unplugged} ones, with a vacant bound state. The energy needed to vacate the bound state determines the frequency of the oscillation. The damping is due to the distinct decay rates: the amplitude of the interference diminishes with time because the unplugged states decay faster than the plugged ones. In summary, the time dependence of the fidelity displays rich physics, {bringing to light features that are hardly discernible in plots of the photoemission rate as a function of frequency}.

From an operational perspective, {the computation of the fidelity is simpler than straightforward applications} of the NRG procedure, such as the computation of thermodynamic properties for the Kondo model. The latter calculation relies on three controllable approximations, two of which arise in constructing the renormalization-group transformation: (i) the logarithmic discretization of the conduction band, followed by an exact Lanczos transformation, and (ii) the infrared truncation of the resulting infinite series.
{The subsequent iterative diagonalization of the resulting Hamiltonian calls for an additional approximation: (iii) the ultraviolet truncation of the Hamiltonian matrix at each iteration}. 

{The ultraviolet truncation introduces inaccuracies in the time-dependent properties}, which calls for a more elaborate procedure \cite{2005AnS196801}. By contrast, in the photoemission problem, the initial and final Hamiltonians are quadratic and dispense with iterative diagonalization and, consequently, with ultraviolet truncation.  {Only approximations (i) and (ii) are required}. Approximation (ii) is not a concern, since a modest computational effort is sufficient to keep the resulting deviations under control. We will therefore {focus our attention} on the inaccuracies resulting from the discretization.

%To gauge such imprecision, we have established two benchmarks. One is the time dependence resulting from brute-force numerical diagonalization of the model Hamiltonian, in which the conduction band is a tight-binding Hamiltonian on a lattice with $2L$ sites. Its numerical diagonalization is straightforward for moderately large $L$, which determines the fidelity up to a time proportional to $L$. The other benchmark is an analytical expression for the time-dependent fidelity that is very accurate, except for short times. 

To gauge such imprecision, we have established two {benchmarks. One} is the time dependence resulting from brute-force numerical diagonalization of the model  {Hamiltonian. The} other benchmark is an analytical expression for the time-dependent fidelity that is very accurate, except for short times. Therefore, the two benchmarks are complementary—the numerical diagonalization yields essentially exact results at short and moderately large times. Starting at moderately large times, the analytical expression is likewise precise. From the two, we can always choose a nearly exact value for the fidelity at any time $t$. We call this set of data the \emph{reference} values. Armed with them, we focus our attention on the renormalization-group procedure.

Our study builds upon the real space formulation known as eNRG \cite{2022FeO075129} instead of Wilson's momentum space approach. This formulation is more flexible as it incorporates three dimensionless parameters rather than two: the \emph{discretization} parameter, a real number $\lambda>1$  that corresponds to Wilson's $\Lambda$; the \emph{twist}, a real number $1\ge \theta\ge -1$ that corresponds to $z$ in NRG parlance \cite{oliveira1994}; and the \emph{offset}, a natural number $\zeta$.

Subtraction of the reference fidelities from the output of an eNRG run with specific $\lambda$, $\theta$, and $\zeta$ determines the absolute deviations, which fluctuate and can be relatively large. Our problem, then, is to purge the calculated fidelities from these artifacts of the discretization.

%These deviations are comparable to those that arise in NRG computations of thermodynamic properties. The pioneering calculations of the impurity magnetic susceptibility for the Kondo and Anderson models encountered irregular thermal dependencies at high temperatures and small unphysical oscillations elsewhere in plots of the susceptibility as a function of temperature \cite{wilson1975,1980KWW1003,1980KWW1044}. We call \emph{transitory} the first type of deviation and \emph{persistent} the second. The transitory deviations are due to operators distinguishing the discretized conduction band Hamiltonian from the corresponding fixed point of the renormalization group transformation. These operators are irrelevant: under this transformation, they decay proportionally to the energy or even faster. The renormalization-group theory predicted that their contribution to the physical properties would be significant at thermal energies of the order of the conduction bandwidth and to die out at lower temperatures---the domain of interest on the $T$ axis. The numerical data confirmed this prediction \cite{1980KWW1003}.

These deviations are comparable to those {encountered in NRG calculations of thermodynamic properties for the Kondo and Anderson models \cite{wilson1975,1980KWW1003,1980KWW1044}, where irregular high-temperature behavior and small unphysical oscillations as a function of temperature were observed}. We call \emph{transitory} the first type of deviation and \emph{persistent} the second. The transitory deviations are due to operators distinguishing the discretized conduction band Hamiltonian from the corresponding fixed point of the renormalization group transformation. These operators are irrelevant: under this transformation, they decay proportionally to the energy or even faster \cite{1980KWW1003}. 

%The renormalization-group theory predicted that their contribution to the physical properties would be significant at thermal energies of the order of the conduction bandwidth and to die out at lower temperatures---the domain of interest on the $T$ axis. The numerical data confirmed this prediction \cite{1980KWW1003}.

Persistent deviations are mathematical analogs of the de Haas-van Alphen oscillations \cite{AsM1976solid}. Just as a magnetic field bunches the continuum of conduction states into Landau levels and produces de Haas–van Alphen oscillations, the logarithmic discretization of the band introduces artificial oscillations in the thermal dependence of thermodynamic properties. The NRG parameter $z$ shifts the logarithmic sequence uniformly and hence shifts the phase of the oscillatory deviations. Averaging the calculated thermodynamical property on a set of uniformly spaced $z$'s is therefore sufficient to remove them \cite{oliveira1994}.

The same two sources of deviations arise in the calculations of the fidelity, with more damaging consequences. The irrelevant operators again give rise to transitory deviations, but their decay is slow because it is controlled by the uncertainty principle, not by the Boltzmann factor. The discretization of the conduction levels introduces persistent fluctuations analogous to the harmonic oscillations in the thermal dependence of the thermodynamic properties. Here, however, the persistent deviations are polychromatic, their Fourier spectrum extending to very low frequencies.

To minimize transitory and persistent deviations, we explore two smoothing algorithms. The first is simple, conceptually and operationally. We let the offset $\zeta$ be a large integer, fix the other eNRG parameters, and compute the fidelity from the eigenvalues and eigenvectors of the resulting Hamiltonian. This procedure yields essentially exact fidelities, free of transitory or persistent deviations, over an interval ranging from $t=0$ to a maximum time $T_\zeta\z$ proportional to $\zeta$.
%In other words, it solves the photoemission problem in the time domain.

%As effective as this procedure is in managing quadratic Hamiltonians, the prospect of applying it to models of correlated impurities coupled to conduction bands comes with a limitation imposed by considerations of computational cost.The effort to diagonalize Hamiltonians of this kind grows exponentially with $\zeta$. Our numerical results indicate that $\zeta$ must be somewhat larger than the maximum time of interest $t_{\mathrm{max}}$ in units of the tight-binding hopping parameter. Therefore, it would be prohibitive to adjust $\tz$ to the time scale of screening in the Kondo problem, for example (This difficulty has surfaced in a recent calculation of spectral densities for the two-channel Kondo model \cite{2025PHvDG}, which then used a more involved NRG construction to derive a discretized Hamiltonian analogous to Eq.~\eqref{eq:69} with $\zeta\le20$). However, this rapid growth does not preclude computations with smaller $\zeta$'s, sufficient to make $\tz$ encompass the decay of the transitory fluctuations. Only the problem of calculating the time-dependent properties for $t>\tz$ remains, which leads us to the second smoothing algorithm. 

{Despite this procedure being effective for quadratic Hamiltonians, its extension to correlated impurity models is limited by the exponential growth of the computational cost with $\zeta$. As a consequence, adjusting $\tz$ to the screening timescale of the Kondo problem becomes prohibitive \footnote{This difficulty appeared, for example, in a recent calculation of spectral densities for the two-channel Kondo model \cite{2025PHvDG}, where a more elaborate NRG construction was employed to derive a discretized Hamiltonian analogous to Eq.~\eqref{eq:69} with $\zeta \le 20$.}. Nevertheless, this rapid growth does not prevent calculations for smaller $\zeta$, which are already sufficient for $\tz$ to encompass the decay of transient fluctuations. The remaining challenge is then the calculation of time-dependent properties for $t>\tz$, motivating the second smoothing algorithm}.

The second algorithm combines averaging on $\zeta$ with numerical integration on $\theta$ \cite{2022FeO075129}---an approach reminiscent of $z$-averaging in the NRG method \cite{oliveira1994}. Our results demonstrate that adequate tuning of the parameters eliminates the persistent fluctuations due to the discretization of the conduction energies.

Our study of the time-dependent photoemission rate, therefore, comprises two complementary eNRG procedures. The first offers essentially exact results over a time interval with an extension limited by considerations of computational cost. The second is unable to deal with the transitory fluctuations, but becomes effective at long times because it washes out the persistent artifacts of the discretization. A combination of the two methods emerges as a promising tool.

\section{Model}
\label{sec:model}
In our description of photoemission from a simple metal, a spinless tight-binding Hamiltonian represents the conduction band. The band being half-filled, it is convenient to consider a lattice with an even number $\sz$ of sites. The model Hamiltonian then reads
\begin{align}
\label{eq:1}
\mathcal{H} = \tau\sum_{\ell=0}^{\sz-1}(c_\ell^\dagger
  c_{\ell+1}+\hc)+K(1-d^\dagger d)c_0^\dagger c_0\z,   
\end{align}
where the first term within parentheses on the right-hand side represents the conduction band,
of width $4\tau$, while the second term models the electrostatic interaction, of strength $K$, between a core state
$d$ and the first site $c_{\ell=0}$ in the tight-binding chain. The constant $K$ is negative to make the
interaction attractive. Notational brevity recommends the definition 
\begin{align}
  \label{eq:36}
  \omtau \equiv \dfrac{\tau}{\hbar}.
\end{align}
%$\omtau \equiv \sfrac{\tau}{\hbar}$.

Initially, with an occupied core level, the last term on the right-hand side of Eq.~\eqref{eq:1} vanishes. At $t=0$, an x-ray photon ejects the core electron to a state outside of the metal. 
%and suddenly creates the attractive core-hole potential $K<0$. 
The many-body state $\ket{\Psi_{t}\z}$ initially coincides with the ground state $\ket{\Omega_0\z}$ of the initial Hamiltonian, that is, the $t<0$ Hamiltonian. For $t\ge 0$, after the sudden photoejection,  the core state $d$ is vacant, so that $\mathcal{H}$ becomes equivalent to the final Hamiltonian
\begin{align}
\label{eq:2}
  \mathcal{H}_K =
  \tau\sum_{\ell=0}^{\sz-1}(c_\ell^\dagger c_{\ell+1}+\hc)+K c_0^\dagger c_0\z.
\end{align}

%\section{Fidelity: definition and short-time behavior}\label{sec:fidel}
\section{Fidelity as a Measure of the Photoemission Current}\label{sec:fidel}

XPS measures the Fourier transform of the photoemission current \cite{1970DoS285,Mahan2010-xj}. In the time domain, the current is proportional to the probability $\abs{\braket{\Omega_0\z}{\Psi_t\z}}^2$ of finding the system in the initial ground state at the time $t$.
Seeing that $\ket{\Psi_0\z}$ coincides with $\ket{\Omega_0\z}$, we henceforth focus our attention on the \emph{fidelity}
\begin{align}\label{eq:3}
\ff(t) \equiv \left|\braket{\Psi_0\z}{\Psi_t\z}\right|^2
\end{align}
of the time-dependent state vector $\ket{\Psi_t\z}$.

\subsection{Short-time behavior}\label{sec:fidel_short-time}
Given the Hamiltonian~\eqref{eq:2}, the following expression determines the fidelity:
\begin{align}
  \label{eq:4}
  \ff(t) =
  \abs{\mel{\Psi_0}{\exp(-i\dfrac{ \mathcal{H}_K t}{\hbar})}{\Psi_0}}^2.
\end{align}

At $t=0$, Eq.~\eqref{eq:4} yields $\ff(0)=1$. To calculate $\ff(t>0)$, we must deal with the time-evolution operator in the matrix element. At short times, a Trotter-Suzuki decomposition leads to a simple, accurate expression. \lno{The straightforward algebra, widely used in numerical procedures (see, for example, \cite{2017Hal,2023BFSpL022003}),} starts by splitting the right-hand side of Eq.~\eqref{eq:2} into the first and second terms:
\begin{align}\label{eq:5}
\mathcal{H}_K\z = \hh_0\z + W_K\z,
\end{align}
where
\begin{align}\label{eq:6}
    W_K\z \equiv {K} c_0^\dagger c_0\z.
\end{align}

Using Eq.~\eqref{eq:5}, we expand the time-evolution operator on the right-hand side of Eq.~\eqref{eq:4} in a Maclaurin series up to third order in the exponent. Rearrangement of the resulting sum then shows that
\begin{align}
  \label{eq:8}
  \mel{\Psi_0}{e^{-i\mathlarger{\frac{\hh_K\z t}{\hbar}}}}{\Psi_0} = &\dfrac{1}{2}
  \mel{\Psi_0}{e^{-i\mathlarger{\frac{\hh_0\z t}{\hbar}}}
  e^{-i\mathlarger{\frac{ W_K t}{\hbar}}}}{\Psi_0}\nonumber\\
  &+\dfrac12\mel{\Psi_0}{e^{-i\mathlarger{\frac{ W_K t}{\hbar}}}e^{-i\mathlarger{\frac{\hh_0\z t}{\hbar}}}}{\Psi_0}\nonumber\\
  &-\dfrac{4 i}{9\pi}\left(\frac{K}{\tau}\right)^2(\omtau t)^3.
\end{align}

At short times $\left(\left(\frac{K}{\tau}\right)^2(\omtau t)^3\ll 1\right)$, we can drop the last term on the right-hand side of Eq. \eqref{eq:8}, and it follows that
\begin{align}
  \label{eq:9}
  \mel{\Psi_0}{e^{-i\mathlarger{\frac{ \hh_K\z t}{\hbar}}}}{\Psi_0} =&
  e^{-i\Gusfrac{ E_0 t}{\hbar}} 
  \mel{\Psi_0}{e^{-i\Gusfrac{ W_Kt}{\hbar}}}{\Psi_0},
  %\nonumber\\
  %&\qquad\left(\left(\dfrac{K}{\tau}\right)^2(\omtau t)^3\ll 1\right),
\end{align}
where $E_0$ is the ground-state energy of the initial Hamiltonian.

We can easily compute the matrix element on the right-hand side of Eq.~\eqref{eq:9}. Inspection of Eq.~(\ref{eq:6}) shows that $W_K$ is a projection operator. All terms in the Maclaurin expansion of $\exp(-i\frac{W_Kt}{\hbar})-1$ are hence proportional to $c_0^\dagger c_0\z$, with the result
\begin{align}
  \label{eq:10}
e^{-i\Gusfrac{W_Kt}{\hbar}}= 
1 + \left(\exp(-i\dfrac{K t}{\hbar})-1\right)c_0^\dagger c_0\z.
\end{align}

The $c_0\z$ orbital is half-filled in the initial state, and so
\begin{align}\label{eq:11}
  \mel{\Psi_0\z}{e^{-i\mathlarger{\frac{\hh_K\z t}{\hbar}}}}{\Psi_0\z} =
  e^{-i \Gusfrac{E_0\z t}{\hbar}}
  \dfrac{1+ e^{-i \Gusfrac{K t}{\hbar}}}{2},
\end{align}
which determines the fidelity in the time interval $
  (\omtau t)^3\ll (\sfrac{\tau}K)^2$:
\begin{align}
  \label{eq:12}
  \ff(t) = \cos^2\left(\dfrac{K t}{2\hbar}\right).
\end{align}

Of special interest is the initial curvature of $\ff$. Differentiation of Eq.~\eqref{eq:4} proves that the curvature is proportional to the
variance $\sigma_H\z$ of the Hamiltonian:
\begin{align}
  \label{eq:13}
  \left.\dv[2]{\ff}{t}\right|_{t=0} = -\dfrac{\sigma_H^2}{\hbar^2},
\end{align}
where
\begin{align}
  \label{eq:14}
  \sigma_H^2 \equiv \mel{\Psi_0\z}{ \mathcal H^2}{\Psi_0}
-{\mel{\Psi_0\z}{\mathcal H}{\Psi_0}}^2. 
\end{align}

From Eqs.~\eqref{eq:12}~and \eqref{eq:13}, it results that
\begin{align}
  \label{eq:15}
  \sigma_H^2 = \dfrac{K^2}{4}.
\end{align}

The standard deviation $\sigma_H\z$ is equal to the quantum speed limit $\nu_{QSL}\z$ \cite{2020FDB+110601}, which therefore grows without bounds as $|{K}| \to\infty$. The right-hand  side of Eq.~\eqref{eq:12} vanishes at $t_K\z\equiv\sfrac{\pi \hbar}{|K|}$. The time domain where Eq.~\eqref{eq:12} is valid shrinks as $|{K}|$ increases. Nonetheless, for sufficiently strong potential, $t_K\z$ enters that domain. We conclude that for $|{K}|\to\infty$, the fidelity vanishes rapidly, in agreement with~\cite{2020FDB+110601}.

If the potential is very weak, with $|{K}|\ll 1$, Eq.~\eqref{eq:12} remains valid well beyond $\omtau t= 1$. However, we are also interested in stronger $K$'s, in which case the Trotter-Suzuki expansion breaks down at short times. The numerical treatment is then the simplest alternative, {starting with the diagonalization of the initial and final Hamiltonians.}

\subsection{Diagonalization of $\hh_0$ and $\hh_K$ }

The quadratic Hamiltonian $\hh_0\z$ can be easily diagonalized, analytically \cite{diniz2025trackingadiabaticitynonequilibriummanybody} or numerically, to yield $\sz$ nondegenerate single-particle eigenvalues. Half of them are positive, and the other half are negative. It is practical to label these energy levels in ascending order, starting with the lowest one. In this notation, the initial Hamiltonian, on its diagonal basis, reads
\begin{align}
  \label{eq:20}
  \hh_0\z = \sum_{n=-L}^{L-1}\varepsilon_n\z a_n^\dagger a_n\z,
\end{align}
where {the energies are}
\begin{align}
  \label{eq:21}
  \varepsilon_{n}\z = 2\tau\sin(\dfrac{\pi}{2L} \left(n+\dfrac12\right))\qquad(n=-L,\ldots,L-1),
\end{align}
{for open boundary conditions.}
%In this sequence, the highest negative and lowest positive eigenvalues are $\varepsilon_{-1}\z$ and $\varepsilon_0\z$, respectively.
In the ground state $\ket{\Psi_0\z}$, the negative levels $\varepsilon_{n}$ are filled. The Fermi level lies at zero energy, halfway between $\varepsilon_{-1}\z$ and $\varepsilon_{0}\z$. 

For large $L$, in the vicinity of the Fermi level, Eq.~(\ref{eq:21}) reduces to the approximate expression
\begin{align}
  \label{eq:22}
  \varepsilon_n\z \approx \dfrac{\tau\pi}{L} \left(n+\dfrac12\right)\qquad(\abs{n}\ll L),
\end{align}
which defines a set of equally spaced levels.

The final Hamiltonian $\hh_K\z$ is particle-hole asymmetric. Nonetheless, it is still quadratic and can be likewise diagonalized \cite{diniz2025trackingadiabaticitynonequilibriummanybody}:
\begin{align}
  \label{eq:23}
  \hh_K\z = \sum_{n=-L}^{L-1}\epsilon_{n}\z b_{n}^\dagger b_{n}\z,
\end{align}
where
\begin{align}\label{eq:24}
  \epsilon_{n}\z &= 2\tau\sin(\dfrac{\pi}{\sz}\left(n+\dfrac12- \dfrac{\delta}{\pi}\right))\nonumber \\
  &~~(n=-(L-1),\ldots, L-1),
\end{align}
with a \emph{phase shift} $\delta$ fixed by the scattering potential \footnote{Reference ~\cite{diniz2025trackingadiabaticitynonequilibriummanybody} derives Eq.~\eqref{eq:25} for the tight-binding model.} and the Fermi-level density of states $\rho = 1/(2\pi\tau)$:
\begin{align}
  \label{eq:25}
  \tan\delta = - 2\pi\rho K = -K/\tau.
\end{align}

For large $L$, close to the Fermi level, Eq.~\eqref{eq:24} takes an approximate form analogous to Eq.~\eqref{eq:22}:
\begin{align}
  \label{eq:26}
  \epsilon_{n}\z \approx \dfrac{\tau\pi}{L}\left(n+\dfrac12 - \frac{\delta}{\pi}\right)\qquad(\abs{n} \ll L).
\end{align}

 Equation~\eqref{eq:26} shows that the splitting $\dl$ between successive single-particle eigenvalues near the Fermi level is the same in the initial and final Hamiltonians:
\begin{align}
  \label{eq:27}
  \dl = \dfrac{\tau\pi}{L}.
\end{align}

The eigenstates of $\mathcal{H}_K\z$ can be written as linear combinations of the eigenstates of $\mathcal{H}_0\z$. Except for the eigenstate $b_{-L}\z$, discussed below, the following expression accurately describes the linear coefficients \footnote{For a derivation of Eq.~\eqref{eq:28}, see~\cite{diniz2025trackingadiabaticitynonequilibriummanybody}}:
\begin{align}\label{eq:28}
  \acomm{a_m^\dagger}{b_n\z} =& \dfrac{\sin(\delta)}{\varepsilon_m\z-\epsilon_n\z} \frac{\dl}{\pi}\nonumber\\
  (m=-L,\ldots, L-1;\  n=& -(L-1), \ldots, L-1).
  \end{align}

The eigenstate $b_{-L}\z$, associated with the lowest eigenvalue $\epsilon_{-L}\z$, is a bound state. For small $|K|$, Eq.~(\ref{eq:24}), with $n=-L$, also describes its energy. However, as the potential becomes more negative and exceeds $|K| = \tau$, the lowest level splits off the bottom of the conduction band. The following equality then describes its energy \footnote{For $\abs{K} > \tau$ and $L>10$, Eq.~\eqref{eq:29} reproduces the numerically-computed bound states with deviations smaller than $0.1\%$.}:
\begin{align}
  \label{eq:29}
  \epsilon_{-L}\z = K + \dfrac{\tau^2}{K}.
  % \epsilon_{-L}\z = K \left(1 + \dfrac{\tau^2}{K^2}\right).
\end{align}

%%%%%%%%%

\subsection{Time dependence of the fidelity}
Two procedures are available to determine the time dependence of the fidelity. One of them deals with the many-body spectrum of the final Hamiltonian and will be called the \emph{many-body} approach, {which is convenient for the analytical derivations}. The other involves only the single-particle eigenstates and eigenvalues of the initial and final Hamiltonians and will hence be dubbed the \emph{single-particle} method, {which provides an efficient computational algorithm}.

\subsubsection{Many-body procedure}
\label{sec:many-body-procedure}
The vector $\ket{\Psi_t}$ \fdp{is} a linear combination of the many-body eigenstates $\ket{r}$ of the Hamiltonian~\eqref{eq:2}:
\begin{align}\label{eq:16}
    \ket{\Psi_t\z} = \sum_r\z \alpha_r\z\exp(-i \dfrac{E_r\z t}{\hbar})\ket{r}, 
 \end{align}
 where $E_r\z$ denotes the eigenvalue associated with $\ket{r}$, and the coefficients $\alpha_r = \braket{r}{\Psi_{0}\z}$, which leads to a simple expression for the fidelity: 
\begin{align}\label{eq:17}
  \ff(t) =
  \abs{\sum_r\z {\abs{\braket{r}{\Psi_0\z}}}^2\exp(-i\dfrac{E_r\z  t}{\hbar})}^2. 
\end{align}

%On the right-hand side of Eq.~(\ref{eq:17}), a sum is multiplied by its complex conjugate. The expansion of the product shows that
The expansion of the product on the right-hand side of Eq.~(\ref{eq:17}) shows that
\begin{align}
  \label{eq:18}
  \ff(t) =
  \sum_{r\ne r'}\z &{\abs{\braket{r}{\Psi_0\z}}}^2 {\abs{\braket{r'}{\Psi_0\z}}}^2\cos(\dfrac{E_r\z-E_{r'}\z}{\hbar}t)\nonumber\\
  &+\sum_r\z\abs{\braket{r}{\Psi_0\z}}^4. \end{align}

At $t=0$, the phases $\sfrac{E_r\z t}{\hbar}$ of the exponentials in the sum on the right-hand side of Eq.~(\ref{eq:17}) vanish. The sum then reduces to the normalization of $\ket{\Psi_0\z}$, and the equality to $\ff(t=0)=1$. As $t$ grows, the eigenstates $\ket{r}$ satisfying $\sfrac{|E_r|\z t}{\hbar} \agt 1$ acquire significantly different phases, and the resulting destructive interferences reduce the fidelity. At very long times, one expects the first term on the right-hand side of Eq.~(\ref{eq:18}) to approach zero and the fidelity to approach the following small constant:
\begin{align}
  \label{eq:19}
  \cres = \sum_r\z\abs{\braket{r}{\Psi_0\z}}^4.
\end{align}

In the continuum limit $L\to\infty$, the overlaps $\braket{r}{\Psi_0\z}$ become infinitesimal, and $\cres\to 0$, a consequence of the Anderson orthogonality catastrophe \cite{1967And1049}. However, for finite lattices, the sum on the right-hand side of Eq.~\eqref{eq:19} is nonzero, and so $\ff(t\to\infty)\ne 0$. The numerical results in the following sections will confirm these expectations.
 
Consider now the practical aspects of computing the fidelity. For small $L$, given the eigenvalues and eigenvectors of $\hh_0\z$ and $\hh_K\z$, {computing the right-hand side of Eq.~(\ref{eq:17}) is straightforward. However, the computational cost rapidly becomes prohibitive as the lattice size grows because the number of contributing many-body eigenstates is $N_r = \binom{\sz}{L}$. In practice, therefore, an alternative numerical approach is preferable.}

%Consider now the practical aspects of computing the fidelity. For small $L$, given the eigenvalues and eigenvectors of $\hh_0\z$ and $\hh_K\z$, it is a simple numerical exercise to compute the right-hand side of Eq.~(\ref{eq:17}) However, the computational cost rapidly becomes prohibitive as the lattice size grows because the number of many-body eigenstates contributing to the sum is
%\begin{align}\label{eq:30}
%N_r = \binom{\sz}{L}.
%\end{align}
%In practice, therefore, an alternative numerical approach is preferable.

\subsubsection{Single-particle method}
\label{sec:single-part-meth}

\lno{For quadratic Hamiltonians} and large $L$, \lno{an alternative approach proves more practical \cite{1969NoD1097,Knap2012}. We} rewrite the projection on the right-hand side of Eq.~\eqref{eq:3} as
\begin{align}\label{eq:31}
\braket{\Psi_0\z}{\Psi_t\z}= \mel{\emptyset}{a_{-1}\z\ldots a_{-L}\z a_{-L}^\dagger(t)\ldots a_{-1}^\dagger(t)}{\emptyset}, 
\end{align}
where $\ket{\emptyset}$ is the vacuum state, and $a_{n}^\dagger(t)$ ($n=-L,\ldots, L-1$) denotes the time evolution of the initial-Hamiltonian eigenstate $a_n^\dagger$, driven by the final-state Hamiltonian, that is,
\begin{align}\label{eq:32}
  a_n^\dagger(t) = e^{-i\hh_K\z t/\hbar} a_n^{\dagger}  e^{+i\hh_K\z t/\hbar},
\end{align}
or after projection onto the basis of the final-Hamiltonian eigenstates $b_{\ell}\z$ ($\ell=-L,\ldots, L-1$),
\begin{align}
  \label{eq:33}
  a_n^\dagger(t) = \sum_{\ell=-L}^{L-1} \acomm{a_n^\dagger}{b_{\ell}\z}e^{-i\epsilon_{\ell}\z t/\hbar}
  b_{\ell}^\dagger.
\end{align}

Wick's theorem then turns Eq.~(\ref{eq:31}) into the expression
\begin{align}
  \label{eq:34}
  \braket{\Psi_0\z}{\Psi_t\z} = \det \underline{\underline{M}}(t), 
\end{align}
where $\underline{\underline{M}}(t)$ is the $L\times L$ matrix with elements
\begin{align}\label{eq:35}
  \underline{\underline{M}}_{m,n}(t) = \sum_{\ell=-L}^{L-1}
  \acomm{a_m\z}{b_\ell^\dagger}e^{-i\epsilon_\ell t/\hbar}\acomm{b_\ell\z}{a_n^\dagger}. 
\end{align}

{To accurately compute the fidelity up to a maximum time $ \bar t$}, we must work with an energy splitting~(\ref{eq:27}) small compared to $\hbar/\bar t$, so that $\exp(-\sfrac{i\Delta_L\z \bar{t}}{\hbar})\approx 1$. The condition, therefore, is $L = \sfrac{\tau \bar t}{(\hbar\eta)}$, where $\eta\pi \ll 1$.

%The computational cost of evaluating the determinant on the right-hand side of Eq.~(\ref{eq:34}) at a single instant $t=\bar t$ grows in proportion to ${L^3}$. To accurately compute the fidelity at that time, we must work with an energy splitting~(\ref{eq:27}) small compared to $\hbar/\bar t$, so that $\exp(-\sfrac{i\Delta_L\z \bar{t}}{\hbar})\approx 1$. The condition, therefore, is $L = \sfrac{\tau \bar t}{(\hbar\eta)}$, where $\eta\pi \ll 1$. So, the computational cost of calculating the fidelity in a single instant $\bar{t}$ increases in proportion to $\bar{t}^3$, and a full calculation of the fidelity in the time interval $0\le t\le t_{\mbox{max}}\z$ requires a computational effort proportional to $t_{\mbox{max}}^4$. Compared with directly using Eq.~\eqref{eq:17}, Eq.~\eqref{eq:34} is overwhelmingly more efficient.

\subsection{Time scales}
\label{sec:time-scales}
The coupling $\tau$ fixes the conduction bandwidth, controls the short-time behavior of the fidelity, and, within the context of numerical calculations, makes it convenient to work with dimensionless times $\omtau t$, where $\omtau \equiv \sfrac{\tau}{\hbar}$.
%\begin{align}
%  \label{eq:36}
%  \omtau \equiv \dfrac{\tau}{\hbar}.
%\end{align}

At long times, with $\omtau t\gg 1$, the characteristic time $T_L\z$, defined by the equality
\begin{align}
  \label{eq:37}
  \omtau T_L\z = 2L
\end{align}
sets the time scale.
Specifically, comparison with Eq.~\eqref{eq:27} shows that the smallest energy in the single-particle spectrum is inversely proportional to the characteristic time; namely, 
\begin{align}\label{eq:38}
\dl =2\pi\hbar/ T_L\z.
\end{align}
Since the energy differences in the spectra of $\hh_0\z$ and $\hh_K\z$ are proportional to $\dl$, the periods of the oscillations that contribute to $\ff(t)$ are multiples of the characteristic time $T_L\z$. In other words, at $t=T_L\z$, the fidelity completes a cycle. It follows that $\ff(T_L\z) = 1$, a finite-size effect without physical equivalence in the continuum. 

At moderately long times, such that $t\ll T_L\z$ while $\omtau t\gg1$, energies of $\order{\dl}$ contribute insignificantly to the many-body energy in the exponent $-iE_r\z t/\hbar$ on the right-hand side of Eq.~\eqref{eq:16} and can be neglected. In this time regime, the fidelity is therefore independent of $T_{L}\z$. However, this changes as $t$ increases. An example with $K=-2\tau$ in Sect.~\ref{sec:large-zeta} (see Fig.~\ref{fig:8}(d)) shows deviations of $0.6\%$ at $t= T_L\z /10$, in comparison to the fidelity calculated at the same instant in the limit $L\to\infty$. 

\section{Numerical Results and Physical Interpretation}\label{sec:numerical-results}

\begin{figure}[hbt!]
    \centering \includegraphics[width=0.99\columnwidth]{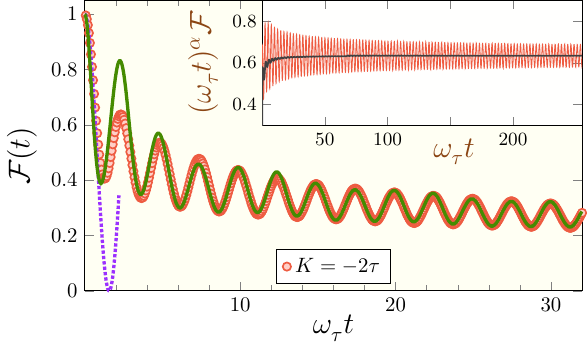}
    \caption[tight-binding:f1]{ \label{fig:1} Decay of the fidelity relative to the initial ground state $\ket{\Psi_0\z}$ following the sudden application of a localized potential $K=-2\tau$. The filled red circles represent numerical results for the tight-binding Hamiltonian~(\ref{eq:2}) with $2L=5\,000$ sites. The dashed purple line interrupted at $\omtau t =2$ represents Eq.~\eqref{eq:12}, valid at short times only, while the solid green line depicts Eq.~\eqref{eq:54}, whose accuracy grows rapidly with time. To evidence alignment with the Doniach-Sunjic power law, the inset shows the numerically computed fidelity multiplied by $(\omtau t)^{\alpha}$, where $\alpha$ is the Doniach-Sunjic exponent, Eq.~\eqref{eq:39}. The solid black line is the running average of the ordinate over each period of oscillation.}
\end{figure}

Figure~\ref{fig:1} shows the fidelity as a function of $\omtau t$, computed from Eq.~(\ref{eq:34}) for the Hamiltonian~(\ref{eq:2}) with $2L=5\,000$ and potential scattering $K=-2\tau$. The red filled circles represent the numerical data computed with Eq.~\eqref{eq:34}. For comparison, the dashed purple line displays Eq.~(\ref{eq:12}), reliable at small times only. Within its validity range, for $\left(\sfrac{K}{\tau}\right)^2(\omtau t)^3 \ll 1$, the dashed curve fits the red circles very well. However, at $\omtau t\approx 1.5$, the plots split apart, the squared cosine on the right-hand side of Eq.~(\ref{eq:12}) dropping to zero and continuing to oscillate with unitary amplitude (not shown) while the circles define a damped oscillatory decay.

The inset shows that the decay of the fidelity follows the power law $F(t)\sim (\omtau t)^{-\alpha}$, with the Doniach-Sunjic exponent~\eqref{eq:39}.

As Ref.~\cite{1970DoS285} pointed out, this decay is a consequence of the Anderson orthogonality catastrophe. The conservation of energy allows the final states to have numerous particle-hole excitations because the potential $K$ lowers the final energy of the ground state relative to the initial one. The uncertainty principle makes the energy $\epsilon_{ph}$ of one such particle-hole pair undistinguishable from zero at small times. More specifically, as long as $\epsilon_{ph}\z<\sfrac{\hbar}{t}$, the pair remains dormant, that is, it makes a time-independent contribution to the projection $\braket{\Psi_t\z}{\Psi_0\z}$. As $t$ increases, the barrier $\sfrac{\hbar}{t}$ is reduced and progressively more particle-hole excitations make time-dependent contributions.  This growing number of excitations gives rise to the Anderson catastrophe, which forces the fidelity to decay following the Doniach-Sunjic law.

%The contribution to the current of all pairs with energies smaller than $\epsilon$ is proportional to $(\rho\epsilon)^{\alpha}$. According to the uncertainty principle, in the instant $t$ only pairs with energy $\epsilon_{ph} < \sfrac{\hbar}{t}$ contribute to the current. Therefore, as $t$ increases, progressively fewer particle-hole excitations contribute, and the fidelity decays following the Doniach-Sunjic law.

The red line in the inset of Fig. \ref{fig:1} represents the fidelity magnified by $(\omtau t)^\alpha$ as a function of $\omtau t$. The black line, which depicts the moving average of the ordinates over each period of oscillation, remains virtually constant after the first few cycles, showing that the long-time behavior of the fidelity conforms to the Doniach-Sunjic power law.
Still, the inset leaves two questions unanswered since it does not explain why the fidelity oscillates or why the amplitude of the oscillations decays, even after multiplication by $(\omtau t)^\alpha$. 

\begin{figure}[hbt!]
  \centering\includegraphics[width=0.6\columnwidth]{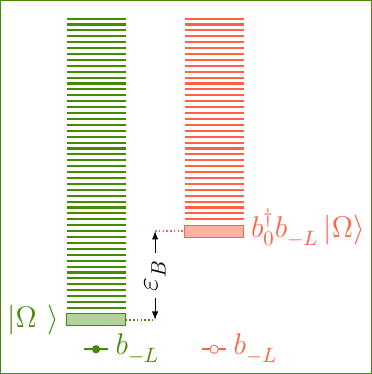}
  \caption[many-body:f2]{\label{fig:2}Classification of the final many-body eigenstates with $L$ electrons. The left (right) column shows the \emph{plugged} (\emph{unplugged}) eigenstates, in which the bound level $b_{-L}\z$ is filled (vacant). The horizontal green (red) thin rectangle represents the ground state (lowest energy unplugged state), and the horizontal green (red) segments above it are many-body excited states.}
\end{figure}

\subsection{Plugged and unplugged many-body eigenstates}\label{sec:Plugg_and_UPlugg_states}
To track the origin of the damped oscillations, it is necessary to examine the eigenstates of the final-state Hamiltonian $\hh_K\z$ that contribute to the fidelity, that is, the many-body eigenstates comprising $L$ occupied levels. The bound state $b_{-L}\z$, with energy $\epsilon_{-L}$, is special among the single-particle eigenstates. Since its occupation $n_{-L}\z\equiv b_{-L}^\dagger b_{-L}$ is conserved, the many-body eigenstates $\ket{r}$ of $\hh_K\z$ can be divided into two classes, depending on whether $n_{-L}\z =1$ or $n_{-L}\z =0$. The depiction in Fig.~\ref{fig:2} suggests the adjectives \emph{plugged} and \emph{unplugged} to label the former and the latter eigenstates, respectively.  The lowest-energy plugged state is the ground state $\ket{\Omega}$ of the final Hamiltonian $\hh_K\z$, with energy $E_0\z$. The lowest unplugged state is $b_0^\dagger b_{-L} \ket{\Omega}$, since $b_{0}\z$ is the first single-particle level with positive energy, and its energy is $E_0\z+\epsilon_B\z$, where $\epsilon_B\z$ is the energy needed to promote an electron from the bound state to the first unoccupied level above the Fermi level, as the double-headed arrow in the figure indicates,
\begin{align}
  \label{eq:40}
  \epsilon_B\z = \dfrac{\dl}2 -\epsilon_{-L}\z.
\end{align}

\subsubsection{One-to-one correspondence between the plugged and unplugged eigenstates}
Let $\mathcal{P}$ denote the set of all the plugged states and $\mathcal{U}$, the set of the unplugged states. There is a one-to-one correspondence between the states in $\mathcal{P}$ and in $\mathcal{U}$. To verify this statement, take any plugged state $\ket{P}\in\mathcal{P}$ and construct a many-body state $\ket{U}$ so that (i) the bound state is vacant and (ii) the occupation of each level $\epsilon_p\z$ ($p=-L+1, \ldots, L-1$) is equal to the occupation of level $\epsilon_{p-1}$ in $\ket{P}$.

\begin{figure}[!ht]
  \centering
  \includegraphics[width=0.68\columnwidth]{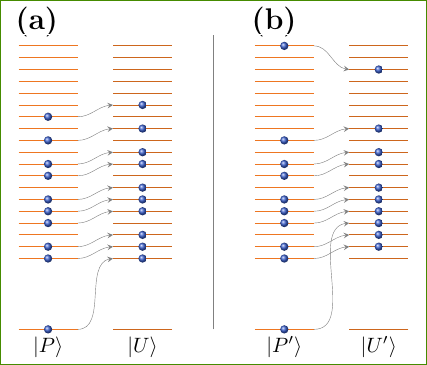}
  \caption[p2u:fig3]{\label{fig:3}Construction of unplugged many-body eigenstates from plugged eigenstates. In each panel, the orange horizontal lines depict single-particle levels, and the blue circles mark the occupied levels. The bound state, split off the bottom of the conduction band, is occupied in the plugged state and vacant in the unplugged one. Panel (a) depicts the standard construction applicable to plugged states in which the highest-energy conduction level is vacant. As the curved arrows indicate, for each occupied level in $\ket{P}$, the construction then fills the next higher-energy level in $\ket{U}$. Panel (b) shows the special construction required when an electron occupies the highest-energy level in $\ket{P}$. As the curved arrows indicate, the occupied levels in $\ket{P'}$ between the bound-state energy $\epsilon_{-L}\z$ and the highest energy $\epsilon_{L-1}\z$ follow the upgrade rule in panel (a). The photoemission then promotes the electron in the bound state to the first vacant level above $\epsilon_{-L}\z$ in $\ket{U'}$, and the electron in level $\epsilon_{L-1}\z$ is demoted so that the energy difference between the unplugged and plugged states satisfies Eq.~(\ref{eq:43}).}
\end{figure}

As the arrows in Fig.~\ref{fig:3}(a) indicate, the energy difference between $\ket{U}$ and $\ket{P}$ is
\begin{align}\label{eq:41}
E_U\z -E_P\z = \epsilon_{-L+1}\z - \epsilon_{-L}\z + (L-1)\dl,
 \end{align}
 and since $(L-1)\dl$ is the energy difference between $\epsilon_{-L+1}\z$ and $\epsilon_0\z$, it follows from Eq.~\eqref{eq:40} that
 \begin{align}\label{eq:42}
  E_U\z - E_P\z = \eb.
\end{align}

One class of plugged states requires special handling. The assignment depicted by the arrows in Fig.~\ref{fig:3}(a) is inapplicable to plugged states such as $\ket{P'}$ in Fig.~\ref{fig:3}(b), in which the highest energy level $\epsilon_{L-1}\z$ is occupied; moreover, the assignment in panel (a) never yields unplugged states such as $\ket{U'}$, in which the level $\epsilon_{-(L-1)}\z$ is vacant. Nonetheless, the special construction described by the arrows in Fig.~\ref{fig:3}(b) associates $\ket{P'}$ with $\ket{U'}$ and guarantees that the energy difference between the two states satisfies the equality
\begin{align}
  \label{eq:43}
  E_{U'}\z - E_{P'}\z = \epsilon_B\z.
\end{align}

Together, the assignments in panels (a) and (b) of Fig.~\ref{fig:3} associate a unique unplugged state with each plugged state. We can reverse the assignments to associate a unique plugged state with each unplugged state. Therefore, there is a one-to-one correspondence between the plugged and unplugged states. Moreover, the energy difference between each unplugged state and the matching plugged state is $\eb$, as Eqs.~\eqref{eq:42}~and \eqref{eq:43} show.

\subsubsection{Contributions from the plugged and unplugged states to the fidelity}
The conclusion that the energy separation between each plugged eigenstate $\ket{P}$ and the corresponding unplugged eigenstate $\ket{U}$ is uniform \fdp{suggests} the following partition of the sum on the right-hand side of Eq.~(\ref{eq:17})
\begin{align}
  \label{eq:44}
  \ff(t)= \abs{\pp_P\z + \exp(-i\omb t)\pp_U\z}^2.
\end{align}
Here, $\omb\equiv \sfrac{\epsilon_B\z}{\hbar}$,
\begin{align}
  \label{eq:45}
  \pp_P\z \equiv \sum_{P}\z \abs{\braket{P}{\Psi_0\z}}^2\exp(-i \dfrac{E_P\z t}{\hbar}),
\end{align}
and
\begin{align}
  \label{eq:46}
  \pp_U\z \equiv \sum_{U}\z \abs{\braket{U}{\Psi_0\z}}^2\exp(-i \dfrac{E_{P(U)}\z t}{\hbar}), 
\end{align}
where the sums span the sets of the plugged eigenstates $\ket{P}$ and the unplugged eigenstates $\ket{U}$, respectively. In Eq.~\eqref{eq:46}, $E_{P(U)}\z \equiv E_U\z - \epsilon_B\z$ denotes the energy of the plugged eigenstate associated with $\ket{U}$.

Although the time-dependent exponentials on the right-hand sides of Eqs.~(\ref{eq:45})~and (\ref{eq:46}) are identical, the projections $\braket{P}{\Psi_0\z}$ and $\braket{U}{\Psi_0\z}$ are different and therefore deserve special attention. Consider a filled level $n$ in the initial ground state $\ket{\Psi_0\z}$ and a filled level $p$ in the plugged state $\ket{P}$ in Fig.~\ref{fig:3}(a). Assuming that neither level lies too far from the Fermi level, so that Eqs.~(\ref{eq:21})~and (\ref{eq:24}) are reliable, the initial energy $\varepsilon_n\z$ and the final energy $\epsilon_p\z$ differ by
\begin{align}
  \label{eq:47}
 \varepsilon_n\z -\epsilon_p\z = \dl\left(n-p+\dfrac{\delta}{\pi}\right).
\end{align}

Next, consider the levels occupied in the unplugged state $\ket{U}$. To each level $p$ in the many-body state $\ket{P}$, the construction in Fig.~\ref{fig:3} associates a filled level $u=p+1$ in $\ket{U}$. The energy difference between the initial level $n$ and the final level $u$ is
$\varepsilon_n\z -\epsilon_u\z$, which can be written as
\begin{align}
   \label{eq:48}
\varepsilon_n\z -\epsilon_u\z   =  \Delta_L\z\left(n-p+ \left(\dfrac{\delta}{\pi}-1\right)\right). 
\end{align}

The difference on the left-hand side of Eq.~(\ref{eq:47}) is the denominator of the fraction on the right-hand side of Eq.~(\ref{eq:28}) and hence determines the projection $\acomm{a_n^\dagger}{b_p\z}$ between the two single-particle eigenstates. The projection $\braket{P}{\Psi_0\z}$, of the plugged many-body state $\ket{P}$ upon the initial state $\ket{\Psi_0\z}$ is the determinant of the $L\times L$ matrix whose elements are the projections $\acomm{a_n\dagger}{b_p\z}$, between the occupied levels $a_n\z$ in $\ket{\Psi_0\z}$ and the occupied levels $b_p\z$ in $\ket{P}$.

Likewise, Eq.~(\ref{eq:48}) determines the projection $\acomm{a_n^\dagger}{b_u\z}$, needed to compute the projection $\braket{U}{\Psi_0\z}$ of the unplugged many-body state $\ket{U}$ upon $\ket{\Psi_0\z}$. The comparison between the right-hand sides of Eqs.~(\ref{eq:47})~and (\ref{eq:48}) shows that the substitution $\delta\to \delta-\pi$ transforms the former into the latter.
  Since $\sin(\delta)$ only changes sign under the same substitution, we can see from Eq.~\eqref{eq:28} that, except for levels close to the band edges,  the construction in Fig.~\ref{fig:3} maps the single-particle projections needed to compute $\braket{P}{\Psi_0\z}$ onto those entering the computation of $\braket{U}{\Psi_0\z}$. In other words, the construction establishes the mapping 
  \begin{align}
    \label{eq:49}
   \abs{\braket{P}{\Psi_0\z}}_{\delta}\mapsto\abs{\braket{U}{\Psi_0\z}}_{\delta-\pi},
  \end{align}
where each subscript is the phase shift needed to calculate the associated many-body projection. 

Inspection of the right-hand sides of Eqs.~\eqref{eq:45}~and \eqref{eq:46} then extends the mapping to the sums $\pp_P\z$ and $\pp_U\z$, namely,
\begin{align}
  \label{eq:50}
  \pp_{P,\delta}\z(t) \mapsto  \pp_{U,(\delta-\pi)}(t).
\end{align}

We only have to compare Eqs.~\eqref{eq:53}~and \eqref{eq:39} to identify the analogous mapping between the Doniach-Sunjic and the Nozières-De Dominicis exponents: 
\begin{align}\label{eq:501}
\beta\mapsto \alpha.
\end{align}

\subsubsection{Interplay between the Doniach-Sunjic and Nozières-De Dominicis components of the fidelity}
The fidelity, given by Eq.~\eqref{eq:44}, results from the interference between two similar photoemission processes with distinct threshold energies, split by $\epsilon_B\z$. The first process involves excitations to the plugged final states, has a lower threshold energy $E_P\z$, and follows the Doniach-Sunjic law:
\begin{align}
  \label{eq:51}
  \abs{\pp_{P,\delta}\z}^2 \propto (\omega_\tau t)^{-\alpha}.
\end{align}

The second one comprises excitations to the unplugged states and has the threshold energy $E_U\z=E_P\z+\epsilon_B\z$. Given the mappings~\eqref{eq:50}~and \eqref{eq:501}, Eq.~\eqref{eq:51} implies that
\begin{align}
  \label{eq:52}
  \abs{\pp_{U,(\delta-\pi)}\z}^2 \propto (\omega_\tau t)^{-\beta}.
\end{align}

To interpret Eq.~\eqref{eq:52} physically, compare the unplugged many-body states to the final states in an XAS experiment. As illustrated by the electron distributions in states $\ket{U}$ and $\ket{U'}$ in Fig.~\ref{fig:3}, each unplugged state contains an electron that has been promoted from the bound state to a single-particle state above the Fermi level. This process is analogous to the excitation of an electron from the core orbital to a vacant conduction level during the absorption experiment. Consequently, the current due to the unplugged states follows a power law with the XAS exponent~\eqref{eq:53}.

In the phase-shift range of interest, $0<\delta< \sfrac{\pi}2$, the exponent $\beta$ always exceeds $\alpha$. Consequently, the contribution from $\pp_{U,(\delta -\pi)\z}$ to the fidelity dies out faster, as does the amplitude of interference between the two processes. The partition of the final eigenstates into plugged and unplugged states thus explains the damping of the oscillations in Fig.~\ref{fig:1}. The algebraic approach in the following section confirms this argument.

\subsection{\label{sec:analytical-results}  Analytical results}

Appendix~\ref{sec:appendix25} treats Eq.~\eqref{eq:44} analytically, based on two approximations that produce accurate results at large times, when the dominant contribution to the fidelity comes from single-particle eigenstates close to the Fermi level: (i) in the initial Hamiltonian ($K=0$) the energy splitting $\dl$ between neighboring levels is assumed to be uniform throughout the conduction band; (ii) in the final Hamiltonian ($K<0$), the conduction-band levels above the bound state are assumed to be uniformly phase-shifted. A closed expression for the time-dependent fidelity is derived:
\begin{align}
  \label{eq:54}
  \ff(t) = \dfrac{1 + \mathcal{R}_t^2 + 2\rt\cos(\omb t+\vartheta_t\z)}{(1+ Q)^2} \mathcal{T}_t^{-\alpha}
\end{align}
where
\begin{align}
  \label{eq:55}
  \tf\equiv   \exp(\int_{0}^{\pi\omtau t}\dfrac{1- \cos(x)}{x}\,\dd x),
\end{align}
\begin{align}
  \label{eq:56}
  \rt \equiv Q\tf^{\mathlarger{-\frac{\beta -\alpha}2}},
\end{align}
\begin{align}
  \label{eq:57}
 \vartheta_t\z \equiv \dfrac{\beta - \alpha}2 \sii(\pi\omtau t),
\end{align}
with $\sii(x)=\mathlarger{\int_0^x du {\frac{\sin (u)}{u}}}$ and
\begin{align}\label{eq:58}
Q \equiv \frac{\fu(0)}{\fp(0)}.
\end{align}

The ratio on the right-hand side of Eq.~\eqref{eq:58} involves two ground-state projections and, therefore, depends only on the phase shift $\delta$. As explained in the following, the numerical computation of $\fp(0)$ and $\fu(0)$ is not demanding. Therefore, throughout this report, we have relied on Eq.~\eqref{eq:58} to determine $Q$ and obtain the fidelity from the analytical expression~\eqref{eq:54}. For example, to get the green curve in Fig.~\ref{fig:1}, we have used $Q=0.17$.

\subsubsection{Individual contributions of the plugged and unplugged many-body states}

As mentioned previously, Eq.~\eqref{eq:54} is an algebraic statement of the conclusions drawn in Sec.~\ref{sec:Plugg_and_UPlugg_states}. To bring this point home, Appendix~\ref{sec:appendix25} also shows that
\begin{align}
  \label{eq:59}
\pp_{P,\delta}\z(t)\equiv \dfrac{1}{1+Q}  {\tf}^{-\mathlarger{\frac{\alpha}{2}}}\exp(-i\frac{\alpha}2\sii(\pi\omtau t)), 
\end{align}
and that the transformation $\delta \to \delta-\pi$ implies $Q\mapsto 1/Q$. Therefore, under this transformation, Eq.~\eqref{eq:59} becomes
\begin{align}
  \label{eq:60}
 \pp_{U,(\delta-\pi)}\z(t)\equiv \dfrac{Q}{1+Q} {\tf}^{-\mathlarger{\frac{\beta}{2}}}
 \exp(-i\frac{\beta}2
\sii(\pi\omtau t)).
\end{align}

For more explicit comparison between Eqs.~\eqref{eq:59}~and \eqref{eq:60}, with the help of Eq.~\eqref{eq:56} [Eq.~\eqref{eq:57}], we eliminate $Q$ from the numerator of the fraction ($\alpha$ from the argument of the exponential) on the right-hand side of Eq.~\eqref{eq:59}. These substitutions show that
\begin{align}\label{eq:61}
   \pp_{U,(\delta-\pi)}\z(t)\equiv \dfrac{\rt e^{-i \vartheta_t\z} }{1+Q} {\tf}^{-\mathlarger{\frac{\alpha}{2}}}\exp(-i\frac{\alpha}2\sii(\pi\omtau t)).
\end{align}

Appendix~\ref{sec:appendix25} derives Eq.~\eqref{eq:61} independently. Equations~ \eqref{eq:60}~and \eqref{eq:61} establish the mapping $\pp_{P,\delta} \mapsto \pp_{U,(\delta - \pi)}$, between the contributions of the unplugged states and the plugged states to the fidelity, and thus ratify the arguments presented in Sec.~\ref{sec:Plugg_and_UPlugg_states}.

\subsection{Analytical versus numerical results}

Returning the focus to the numerical results, the solid green line in Fig.~\ref{fig:1} represents Eq.~\eqref{eq:54}. For $\omtau t > 15$, the agreement with the numerical data is excellent. The discrepancies at smaller times, most prominent near $\omtau t= 2$, are due to the approximations supporting the derivation of the analytical result, which become accurate for $\omtau t\gg1$.
Under this condition, Eq.~\eqref{eq:55} admits simplification. Evaluation of the definite integral on the right-hand side yields
\begin{align}
  \label{eq:62}
  \tf(t) = \exp(\gamma + \log(\pi\omtau t) - \ci(\pi\omtau t)),
\end{align}
where $\gamma$ is the Euler-Mascheroni constant and the cosine integral function is $\ci(x)=-\mathlarger{\int_x^\infty du \frac{\cos (u)}{u}}$.

For large values of its argument, the cosine integral on the right-hand side is $\order{(\omtau t)^{-1}}$, and we can drop it. Under this approximation, Eq.~(\ref{eq:62}) reads
\begin{align}
  \label{eq:63}
  \tf = e^\gamma\pi\omtau t,
\end{align}
which reduces Eq.~\eqref{eq:54} to the expression
\begin{align}
  \label{eq:64}
  \ff(t) = \dfrac{\left(\pi e^\gamma \right)^{-\alpha}}{(1+Q)^2}\left(1 + \mathcal{R}_t^2 + 2\rt\cos(\omb t+\vartheta_t\z )\right) (\omtau t)^{-\alpha}. 
\end{align}

Equation~\eqref{eq:64} explains the damping in the inset of Fig.~\ref{fig:1}, as $\mathcal{R}_t$ decays with time. Division of the fidelity by the last factor on its right-hand side yields the ordinate of the inset plot:
\begin{align}
  \label{eq:65}
  (\omtau t)^{\alpha}\ff(t) = \dfrac{\left(\pi e^\gamma \right)^{-\alpha}}{(1+Q)^2}\left(1 + \mathcal{R}_t^2 + 2\rt\cos(\omb t+\vartheta_t\z)\right).
\end{align}

\subsection{Physical interpretation}

The trinomial within parentheses on the right-hand side of Eq.~\eqref{eq:65} is the law-of-cosines sum of two vectors with magnitudes unity and $\rt$ whose phases differ by $(\omb t+\vartheta_t\z)$. The function $\vartheta_t\z$ quickly approaches $\left[\frac{\pi}{4}\left(\beta-\alpha\right)\right]$ as $\omtau t$ increases; therefore, the sum oscillates with frequency $\omb$ and amplitude $2\rt$. From the definition~\eqref{eq:56}~and Eq.~\eqref{eq:63} we can see that $\rt$ vanishes as $\omtau t\to\infty$: the decay of $\rt$ abates the oscillation.

More physical considerations lead to the same conclusion. $\pp_{P,\delta}$ is the contribution of the plugged states to the fidelity, and so $\abs{\pp_{P,\delta}}^2$ follows the Doniach-Sunjic power law \cite{1970DoS285}. Likewise, $\pp_{U,(\delta-\pi)}$ is the contribution of the unplugged states, and $\abs{\pp_{U,(\delta-\pi)}}^2$ follows the Nozières-De Dominicis power law \cite{1969NoD1097}. The energies of the plugged and unplugged states differing by $\eb$, hence the phases of $\pp_{P,\delta}(t)$ and $\pp_{U,(\delta-\pi)}\z(t)$ differ by $\omb t$. Therefore, instead of adding up to $\abs{\pp_{P,\delta}}^2+\abs{\pp_{U,(\delta-\pi)}}^2$, the two contributions interfere. The interference gives rise to the oscillations in Fig.~\ref{fig:1}. The more rapid decay (Nozières-De Dominicis-like) of $\pp_{U,(\delta-\pi)}$ relative to the (Doniach-Sunjic-like) decay of $\pp_{P,\delta}$ accounts for the damping of the oscillations.

To reinforce this interpretation with numerical evidence, consider now the individual contributions $\ffp(t) \equiv |\pp_P(t)|^2$ and $\ffu(t) \equiv |\pp_U(t)|^2$, of the plugged and unplugged eigenstates to the fidelity. To determine them, note that the operator $b_{-L}\z b_{-L}^\dagger$ projects $\ket{\Psi_t}$ on the subset of unplugged eigenstates; accordingly, the following expression determines the unplugged component:
\begin{align}
  \label{eq:66}
  \ff_U(t)\z = {\abs{\mel{\Psi_0\z}{b_{-L}\z b_{-L}^\dagger}{\Psi_t}}}^2.
\end{align}

The following equality expresses the matrix element on the right-hand side of Eq.~\eqref{eq:66}:
\begin{align}
  \label{eq:67}
  &\mel{\Psi_0\z}{b_{-L}\z b_{-L}^\dagger}{\Psi_t}=\nonumber\\
  &\qquad\mel{\emptyset}{a_{-1}\ldots a_{-L} b_{-L}\z b_{-L}^\dagger a_{-L}^\dagger(t)\ldots a_{-1}^\dagger(t)}{\emptyset}. 
\end{align}

Analogously to Eq.~\eqref{eq:31}, Eq.~\eqref{eq:67} can be evaluated using the procedure in Sec.~\ref{sec:single-part-meth}.

Since $b_{-L}^\dagger b_{-L}$ projects $\ket{\Psi_t}$ on the subset of plugged eigenstates, the following expression determines the plugged component of the fidelity:
\begin{align}
  \label{eq:68}
  \ffp(t)\z = {\abs{\mel{\Psi_0\z}{b_{-L}^\dagger b_{-L}\z}{\Psi_t}}}^2, 
\end{align}

The identity $b_{-L}^\dagger b_{-L}= 1- b_{-L}\z b_{-L}^\dagger$ shows that the matrix element in Eq.~\eqref{eq:66} is the difference between the right-hand sides of Eqs.~\eqref{eq:31}~and \eqref{eq:61} and allows calculation of $\ffp(t)$ without additional computational effort.

\begin{figure}[!ht]
  \centering
\includegraphics[width=0.95\columnwidth]{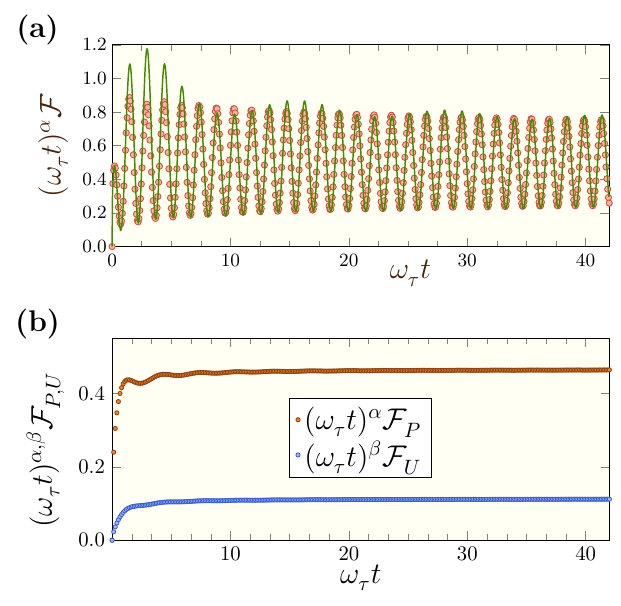}
  \caption[fpu:f4]{\label{fig:4} (a) Comparison between analytical and numerical results for the fidelity. The filled orange circles are the numerical data resulting from straightforward diagonalization of the Hamiltonian~\eqref{eq:1} for $K=-5\tau$ and $2L=5\,000$. The solid green curve represents Eq.~\eqref{eq:65}. For viewing convenience, the analytical and numerical results have been divided by the Doniach-Sunjic law, with $\alpha$ defined by Eq.~\eqref{eq:39}. (b) Absolute values squared of the plugged and unplugged contributions to the fidelity, defined by Eqs.~\eqref{eq:45}~and \eqref{eq:46}, respectively, computed numerically from Eq.~\eqref{eq:67}~and \eqref{eq:68}. The plugged and unplugged contributions $\ffp$ and $\ffu$, represented by the filled orange and blue circles, are divided by the Doniach-Sunjic and Nozières-De Dominicis asymptotic expressions, with exponents defined by Eqs.~\eqref{eq:39}~and \eqref{eq:53}, respectively.}
\end{figure}

\begin{figure}[!ht]
  \centering
\includegraphics[width=1.0\columnwidth]{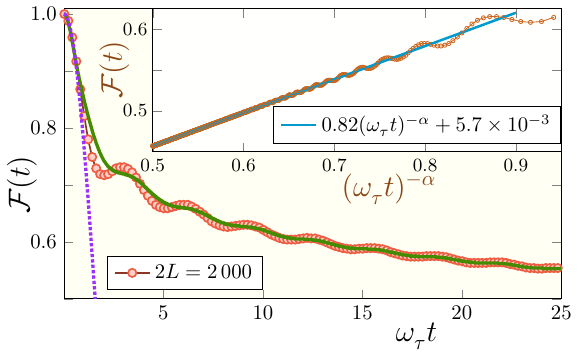}
  \caption[fk1:f5]{\label{fig:5}Fidelity as a function of time for the tight-binding Hamiltonian with 2\,000 sites, and attractive potential $K=-\tau$ (phase shift $\delta=\sfrac{\pi}{4}$). The filled orange circles, dotted purple line, and solid green line represent the numerical data, Eq.~\eqref{eq:12}, and Eq.~\eqref{eq:54}, respectively. The inset displays the fidelity as a function of $(\omega t)^{-\alpha\z}$ to show adherence to the Doniach-Sunjic law as the contribution from the unplugged states vanishes. The straight cyan line is a least-squares regression showing that the fidelity approaches the residual constant $\cres=\num{5.7e-3}$ as $\omtau t\to \infty$, in agreement with Eqs.~\eqref{eq:18}~and \eqref{eq:19}.}
\end{figure}

Figure~\ref{fig:4}(a) displays the time dependence of the fidelity for $K=-5\tau$. The filled orange circles and the solid green line show the numerical and analytical results from Eq. \eqref{eq:65}, respectively, for $\ff(t)$ multiplied by $(\omtau t)^\alpha$. Similarly to Fig.~\ref{fig:1}, the numerical data are remarkably close to the results of Eq.~\eqref{eq:65} for $\omtau t>20$. The phase shift $\delta=0.437\pi$, close to $\sfrac{\pi}{2}$, reduces the exponent on the right-hand side of Eq.~\eqref{eq:56}, which controls the damping of the oscillation. Therefore, in contrast to Fig.~\ref{fig:1} and striking contrast to Fig.~\ref{fig:5} below, the damping is barely perceptible here.

Panel (b) displays the plugged and unplugged components of the fidelity, represented by the orange and blue circles, respectively. The filled orange circles represent $\ffp(t) = \abs{\fp(t)}^2$ amplified by the factor $(\omega t)^\alpha$ for $K=-5\tau$; the rapid convergence to a constant verifies the conformity to the Doniach-Sunjic law. Similarly, the filled blue circles verify that $\ffu(t) = \abs{\fu(t)}^2$ follows the Nozières-De Dominicis law at large times.  The contrast between the vigorous oscillations in panel~(a) and the dull behavior in panel (b) is manifest evidence of the interference between the transitions to plugged and unplugged eigenstates.

Figure~\ref{fig:5} shows the time dependence of the fidelity for a weaker potential $K=-\tau$. The corresponding phase shift in Eq.~\eqref{eq:25} is $\delta=\sfrac{\pi}{4}$. As in Fig.~\ref{fig:1}, only at very short times is there agreement between the filled orange circles and the dashed purple line, which represent the numerical results and Eq.~\eqref{eq:12}, respectively. At long times, Eq.~\eqref{eq:54} describes the fidelity well, as the solid green line indicates. The relatively small phase shift prevents transitions to the unplugged eigenstates and reduces the ratio $Q$ in Eq.~\eqref{eq:58}. Since $\rt$ is proportional to $Q$, the amplitude $2\mathcal{R}_t\z$ of the oscillations becomes much smaller than in Fig.~\ref{fig:1}. Furthermore, while the exponent $-(1-\sfrac{2\delta}{\pi})$ on the right-hand side of Eq.~\eqref{eq:56} was -0.295 for $K=-2\tau$, it is now \sfrac{-1}{2}, which accounts for the rapid dampening of the oscillations.

The inset plots the numerical data as a function of $(\omtau \tau)^{-\alpha}$ to exhibit the behavior over a wider time interval. The oscillations die out as $\omtau t$ grows and, as expected from Eq.~\eqref{eq:18}, the fidelity approaches a small constant. That constant subtracted, the agreement between the filled circles and the Doniach-Sunjic law becomes impeccable at large $\omtau t$, as the straight cyan line shows.

Now that the physical oscillations observed in the photocurrent are well understood, we can turn our attention to reproducing the results discussed in this section using the eNRG method. The motivation here is methodological: to develop an effective smoothing procedure capable of performing time-dependent calculations for non-quadratic Hamiltonians.

\begin{figure}[!ht]
    \centering
    \includegraphics[width=1.0\linewidth]{Dicussion/eNRG_scheme.pdf}
    \caption{The eNRG procedure starts from a 1D tight-binding chain (top, blue circles) and transforms it into a Wilson-like chain (bottom, red circles). The first $\zeta$ sites are copied unchanged from the original chain, and define the off-set region. The remaining sites are grouped through a renormalization procedure, combining $\lambda^n$ sites $(n=0,1,\ldots)$, with $\lambda>1$, into a single operator $f_n$ (red circles), yielding renormalized hopping $\tau_n=\tau\lambda^{\theta-n-1/2}$, where $\theta$ is the twist parameter.}
    \label{fig:eNRG}
\end{figure}

\section{Renormalization-group approach}\label{sec:enrg}

Three integers $\zeta \ge 0$, $\lambda>1$, and $\theta \equiv \log(w)/\log(\lambda)$ ($w\in \mathbb{N}$) parameterize the construction of the eNRG, which divides the $L\to\infty$ tight-binding lattice into an infinite chain of cells, or intervals. The first $\zeta$ intervals $\cl_{\ell}^\zeta$ ($\ell =0,1,\ldots,\zeta$) contain a single site each. The following interval $\cl_0\z$ starts a succession $\cl_n\z$ ($n=0, 1,\ldots$), each member of which encompasses $w\lambda^{n}$ sites.  For each cell $\cl_n\z$, a linear combination of the pertinent operators $c_\ell\z$ then defines a normalized Fermi operator $f_n\z$ \cite{2022FeO075129}. {Figure \ref{fig:eNRG} schematically illustrates this construction}. Although the discretization parameters $\lambda$ and $w$ were defined as integers, Eqs.~\eqref{eq:69}~and \eqref{eq:70} can be rigorously extended to real $\lambda \ge 1$ and $\theta \in [1,-1]$ \cite{2022FeO075129}.

\subsection{Discretized Hamiltonian}
\label{sec:disc}

Along with the $c_\ell\z$ ($\ell=0,1,\ldots,\zeta$), the $f_n\z$ ($n=0,1,\ldots$) form a basis, upon which we project the model Hamiltonian. The projection brings Eq.~\eqref{eq:2} to the form \cite{2022FeO075129}
\begin{align}\label{eq:69}
    \hh_K^{\lambda, \zeta, \theta} =& \left(\tau\sum_{\ell=0}^{\zeta-1} c^{\dagger}_\ell c_{\ell+1}
    +\tau c^\dagger_\zeta f_0\z +\sum_{n=0}^\infty \tau_n f_n^\dagger
  f_{n+1} + \hc\right)\nonumber\\
  &+K c_0^\dagger c_0\z,
\end{align}
where
\begin{align}\label{eq:70}
    \tau_n\z = \tau\lambda^{\mathlarger{-n-\theta-\frac12}}.
\end{align}

The coefficients $\tau_n$ ($n=0,1,\ldots$) in Eq.~\eqref{eq:70} decay exponentially with $n$. To compute the fidelity at time $t$, therefore, it is safe to truncate the infinite series on the right-hand side of Eq.~\eqref{eq:70} at $n=\nn$, where $\nn$ is the smallest integer such that
\begin{align}\label{eq:72}
\dfrac{\dn\z t}{\hbar} < \eta,
\end{align}
where 
\begin{align}\label{eq:74}
    \dn\equiv \tau \lambda^{-\nn-1/2},
\end{align}
is the smallest coefficient $\tau_{n=\nn}\z$ for $\theta=0$. $\eta \ll 1$ is a fixed dimensionless parameter that controls the accuracy of the approximation.

Given $t$, the following equality then defines the truncated scaled Hamiltonian $\hh_K^\nn$:
\begin{align}\label{eq:73}
   {\dn}{\hh_K^\nn} =& \left(\sum_{\ell=0}^{\zeta-1} \tau c^{\dagger}_\ell c_{\ell+1}
    + \tau c^\dagger_\zeta f_0\z +\sum_{n=0}^{\nn-1} \tau_n f_n^\dagger
  f_{n+1} + \hc\right)\nonumber\\
  &+K c_0^\dagger c_0.
\end{align}
This scaling ensures that the spectrum of $\hh_K^\nn$ consists of positive eigenvalues ranging from unity to $2\lambda^{{\nn+1}/2}$ and negative eigenvalues ranging from $-1$ to $-2\lambda^{\nn+1/2}$.

%{It is important to emphasize that the Hamiltonian $\hh_K^\nn$ is quadratic (noninteracting) and therefore does not require the iterative diagonalization procedure used in standard NRG, allowing for efficient diagonalization within the one-particle framework.}

%Given the Hamiltonian~\eqref{eq:73}, two alternative approaches can calculate the fidelity $\ff_{\zeta,\theta}\z(t)$: the many-body approach described in Sec.~\ref{sec:many-body-procedure}  or the more efficient single-particle procedure in Sec.~\ref{sec:single-part-meth}. Either alternative calls for diagonalization of the square matrix of dimension $\nn+\zeta+1$ representing the Hamiltonian $\hh_K^\nn$. Condition~\eqref{eq:72}, with $\tau_\nn\z$ given by Eq.~\eqref{eq:70}, shows that the truncation number $\nn$ grows logarithmically with $t$, a dependence that contrasts with the linear growth of the tight-binding lattice in Sec.~\ref{sec:fidel}. The computational cost of an eNRG run covering a given time interval is, therefore, substantially smaller than that of a tight-binding computation of $\ff(t)$ in the same time range. Section~\ref{sec:numerical-results} presents examples.

{Given the Hamiltonian~\eqref{eq:73}, calculating the fidelity $\ff_{\zeta,\theta}\z(t)$ requires first} diagonalizing the square matrix of dimension $\nn+\zeta+1$ representing the Hamiltonian $\hh_K^\nn$. {It is important to emphasize that this Hamiltonian is quadratic (noninteracting) and therefore does not require the iterative diagonalization procedure of standard NRG, allowing for efficient treatment within the  single-particle procedure in Sec.~\ref{sec:single-part-meth}}. Condition~\eqref{eq:72}, with $\tau_\nn\z$ given by Eq.~\eqref{eq:70}, shows that the truncation number $\nn$ grows logarithmically with $t$, a dependence that contrasts with the linear growth of the tight-binding lattice in Sec.~\ref{sec:fidel}. The computational cost of an eNRG run covering a given time interval is, therefore, substantially smaller than that of a tight-binding computation of $\ff(t)$ in the same time range. Section~\ref{sec:numerical-results} presents examples.

\subsubsection{Single-particle spectrum}
\label{sec:single}
With $K=0$, the right-hand side of Eq.~\eqref{eq:73} is particle-hole symmetric, and so is the resulting single-particle spectrum. Given $\zeta$, it is convenient to choose $\nn$ of the opposite parity so that $\nn+\zeta+1$ is an even number, henceforth denoted $2M$. Of the single-particle eigenvalues, $M$ will be positive and $M$ negative. The same partition results for $K\ne0$, and the following approximate expressions describe the single-particle eigenvalues \cite{2022FeO075129}:
\begin{align}
    \label{eq:76}
    \erg{j}^\pm(K) = \pm\tau \lambda^{\mathlarger{2(j-1)-\theta\pm\frac{2\delta}{\pi}}}
    \qquad\left(j=1,2,\ldots, M\right),
\end{align}
with the phase shift $\delta$ given by Eq.~\eqref{eq:25}.

Equation~\eqref{eq:76} is accurate for the eigenvalues in the range $\dn\ll\dn\abs{\erg{j}\z} \ll \tau$. The inaccuracy grows with $j$, and deviations of a few percent become visible close to the upper limit. For negative $K$, as $\abs{K}$ grows past $2\tau$, the lowest eigenvalue $\erg{-M}\z$ splits off the bottom of the conduction band. For $K< -2\tau$,
\begin{align}\label{eq:77}
  \mathcal{D}_\nn\erg{-M}\z \approx K+\dfrac{\tau^2}{K}.
\end{align}

We will associate the Fermi operators $g_{j\pm}\z$ and $h_{j\pm}\z$ ($j=1,2,\ldots,M$) with the single-particle eigenstates corresponding to the eigenvalues $\erg{j}^\pm(K)$ and $\erg{j}^\pm(0)$  of the final and initial Hamiltonians, respectively.

Since Eq.~\eqref{eq:69} reduces to the tight-binding Hamiltonian~\eqref{eq:2} for $\lambda=1$, to justify the discretization, one must show that the physical properties resulting from the spectrum of \eqref{eq:69} converge to the continuum limit as $\lambda\to1$.

\subsubsection{Computation of the fidelity}\label{sec:numRes}
The analysis leading from Eq.~(\ref{eq:31}) to Eq.~(\ref{eq:35}) can be repeated, with the discretized Hamiltonian~(\ref{eq:73}) substituted for the tight-binding Hamiltonian~\eqref{eq:23}. An expression analogous to Eq.~(\ref{eq:34}) then determines the projection of the time-dependent state on the initial state:
\begin{align}\label{eq:78}
  \braket{\Psi_t\z}{\Psi_0\z} = \det \underline{\underline{M}}^\nn(t), 
\end{align}
where the matrix elements of the $M\cross M$ matrix $\underline{\underline{M}}^\nn(t)$ are
\begin{align}
  \label{eq:79}
  \underline{\underline{M}}_{m,n}^\nn(t) = \sum_p\z\acomm{h_m}{g_p^\dagger}e^{-i\epsilon_p^{\pm} t/\hbar}\acomm{g_p}{h_n^\dagger}.
\end{align}

\subsection{Numerical procedure}
\label{sec:numerical-results-nrg}

  Numerical diagonalization of the initial and the final Hamiltonians $\hh_0^\nn$ and $\hh_K^\nn$ for a specific set of eNRG parameters $\lambda$, $\zeta$, and $\theta$ and adequately chosen truncation number $\nn$, followed by substitution of the right-hand side of Eq.~\eqref{eq:78} for the projection on the right-hand side of Eq.~\eqref{eq:3} yields the \emph{raw} fidelity
  \begin{align}
    \label{eq:80}
    \ff_{\zeta, \theta}\z(t) = {\abs{\det {\underline{\underline{M}}^\nn(t)}}}^2.
  \end{align}

  The eNRG method inherits a trait of the NRG procedure and intersperses plots of physical properties with non-physical oscillations. Thermodynamic properties computed with $\lambda>1$ display sinusoidal functions of the temperature added to the exact thermal dependence, that is, the thermal dependence that a calculation with $\lambda=1$ would produce. In this case, averaging two successive $\zeta$'s, and a few $\theta$'s in the interval $-1\le \theta \le 1$ is enough to eliminate these artifacts \cite{2022FeO075129}.

  The computation of the time-dependent fidelity poses a more complex problem because it involves not one but two discrete spectra, those of the initial and final Hamiltonians. Instead of the uniform energy difference between the plugged and unplugged many-body excitations in Fig.~\ref{fig:2}, we find a variety of splittings, the dispersion of which increases with the discretization parameter $\lambda$. The resulting fidelity is the sum of several artificially distinct frequencies. Gone is the harmony of the damped oscillations in Figs.~\ref{fig:1}, \ref{fig:4}~and \ref{fig:5}.

\subsubsection{Small $\zeta$ and fixed $\theta$}

\begin{figure}[!ht]
  \centering 
  \includegraphics[width=0.99\columnwidth]{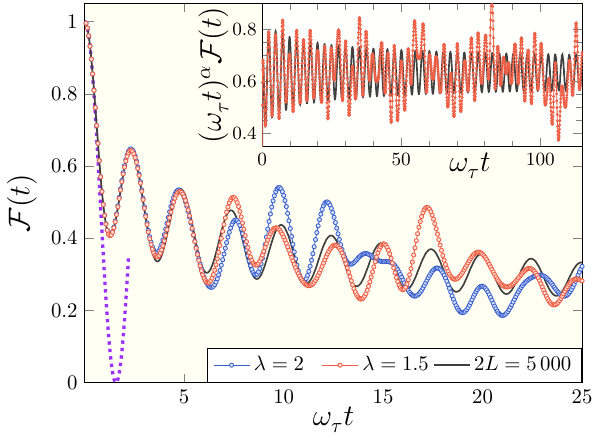}
  \caption[fenrg:f6]{\label{fig:6} eNRG results for the raw fidelity, given by Eq.~\eqref{eq:80}, for $K=-2\tau$ as a function of time scaled by $\omega_\tau\z\equiv \sfrac{\tau}{\hbar}$. The filled orange and blue circles represent the time dependence of the raw fidelity for $\lambda=1.5$ and $\lambda = 2$, respectively, both computed with $\zeta=3$ and $\theta=0$. The dashed purple line is Eq.~\eqref{eq:12}, accurate at short times only. The solid black line is the fidelity for the tight-binding Hamiltonian~\eqref{eq:23} with $2L=5\,000$ lattice sites. The inset compares the data for $\lambda=1.5$ averaged over $\zeta\in \{3,4\}$ (filled orange circles) with the fidelity for the tight-binding Hamiltonian (solid black line). The orange circles show that, divided by the Doniach-Sunjic power law $(\omega_\tau t)^{-\alpha}$, the raw fidelity fluctuates about a constant at long times.}
\end{figure}

As an illustration, Fig.~\ref{fig:6} shows the fidelity resulting from Eq.~\eqref{eq:76} for $\lambda=1.5$ (filled orange circles) and $\lambda = 2$ (filled blue circles), with $K=-2\tau$, $\zeta=3$ and $\theta=0$ in both cases. We have truncated the discretized Hamiltonians at $\nn=34$ ($\lambda=1.5$) and $20$ ($\lambda=2$), respectively, to satisfy \eqref{eq:72} with $\eta=10^{-3}$  up to $\omega_\tau t =1\,000$. We have carried out all the computations in this section with truncation numbers sufficiently large to satisfy that inequality up to the same maximum time. For comparison, the solid black curve shows the essentially exact fidelities calculated from the spectrum of the tight-binding Hamiltonian with $2L=5\,000$ lattice sites.

Like the solid curve, the eNRG curves oscillate with frequency $\omb$. However, undulations with substantially smaller frequencies modulate the plots. These are beat frequencies due to the aforementioned nonuniform energy splittings in the single-particle spectrum~\eqref{eq:76} of the truncated Hamiltonian $\hh_K^\nn$. The unevenness increases with $\lambda$. For this reason, beats with different frequencies and amplitudes distinguish the orange curve from the blue one, with the latter deviating more than the former from the black plot.

The inset compares the $2L=5\,000$ data with the fidelity computed with $\lambda=1.5$ and $\theta=0$, and averaged over $\zeta\in\{3,4\}$. The two fidelities are multiplied by $(\omega_\tau\z \tau)^{-\alpha}$, with the exponent~\eqref{eq:38}, to verify compliance with the Doniach-Sunjic power law \cite{1970DoS285}. Comparison with the orange circles in the main plot shows that averaging on $\zeta$ dampens the deviations but does not eliminate them. Similar irregularities remain after averaging longer sequences, with $\zeta\in \{3,4, 5,6\}$ (not shown), for example. 

In any case, the irregular beats are persistent. With $\zeta\in\{3,4\}$, for example, the pattern in the main plot is repeated up to the longest time studied $\omtau t=10\,000$ (not shown). 

At small times, the eNRG procedure is accurate. In the interval $0\le \omtau t <5$, which extends well beyond the validity of the dotted purple line representing Eq.~\eqref{eq:12} in the main plot, the circles follow the black curve. Outside that range, the orange and blue curves deviate from each other, become irregular, and wobble around the $2L=5\,000$ curve.

\subsubsection{Time scale $\tz$}
\label{sec:large-zeta}

Why are the computed fidelities independent of $\lambda$ at small times? This question prompts us to consider the time scales in the eNRG Hamiltonian.
Section~\ref{sec:time-scales} associated $T_L\z$ with the smallest energy in the single-particle spectrum of the tight-binding Hamiltonian. The analogous energy in the eNRG approach is $\dn$, which defines the time scale $T_\nn\z \equiv \sfrac{\hbar}{\dn}$. At times $t\ll T_\nn\z$, the energy scales close to $\dn$ will contribute insignificantly to the time-evolution operator $\exp(-i\dn\hh_K^\nn t)$; under this condition, $\ff(t)$ will be independent of $\nn$, although it will generally vary with $\lambda$, $\theta$ and $\zeta$. 

However, the latter parameter introduces an exception. The first term within parentheses on the right-hand side of Eq.~\eqref{eq:69} is a segment of the eNRG chain equivalent to a tight-binding lattice of size $\zeta$. Analogy with the discussion of Eq.~\eqref{eq:37} defines a second characteristic time $\tz$, through the equality  
\begin{align}
    \label{eq:81}
    \omtau\tz \equiv \zeta.
\end{align}

In the interval $0\le t<\tz$, the properties of the eNRG Hamiltonian are independent of $\lambda$ and $\theta$, because the energy scales in that segment of the chain, which is not affected by discretization, governs their calculation. The independence of $\lambda$ implies that the fidelities in this interval accurately represent the continuum limit. 

The results in Fig.~\ref{fig:6} were obtained with $\zeta =3$, which corresponds to $\omega \tz= 3$. Close inspection shows that the orange and blue circles become distinguishable from each other and the black line at $\omtau t\approx 3$, that is, $t=T_{\zeta=3}\z$. The good agreement at shorter times confirms that the eNRG data are reliable over the interval defined by $\tz$ and recommends extension to larger $\zeta$.

\subsection{Large $\zeta$ procedure}\label{sec:large_zeta}

\begin{figure}[!ht]
    \centering\includegraphics[width=0.85\columnwidth]{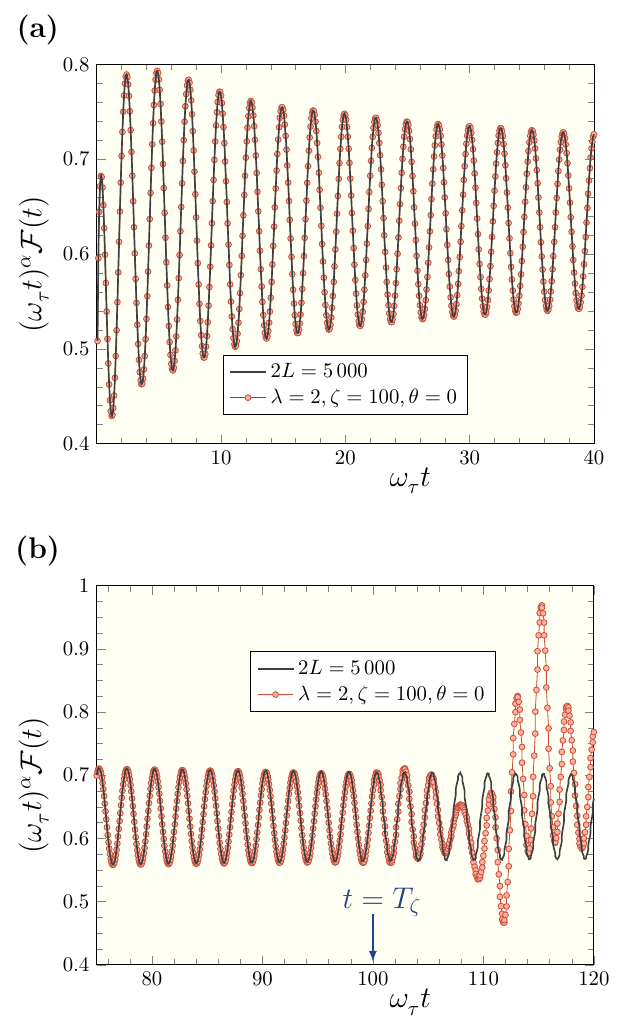}
    \caption[zeta100:f7]{\label{fig:7} Comparison between the time-dependent fidelities for the eNRG Hamiltonian with the displayed parameters and the tight-binding Hamiltonian with $ 2L=5\,000$. In the two panels, the filled orange circles show the eNRG data computed with $\nn =19$. The solid black curve is the tight-binding fidelity. For improved resolution, we have scaled both curves by the Doniach-Sunjic power $(\omega_\tau t)^\alpha$. Panel (a), restricted to short times, shows perfect agreement between the orange circles and the black line.  Panel (b) covers the vicinity of the characteristic eNRG time $T_\zeta\z$, identified by the vertical arrow pointing to the horizontal axis. While flawless for $t< T_\zeta\z$, the agreement deteriorates rapidly as $t$ grows past the characteristic time.}
  \end{figure}

Figure~\ref{fig:7} displays $\ff(t)$ calculated with $\zeta=100$. The two panels compare the raw fidelity computed with $\zeta=100$, $\lambda=2$, and $\theta=0$ (orange circles) with the fidelity for the tight-binding Hamiltonian with $2L=5\ 000$ lattice sites (solid black line). There is excellent agreement until $t \approx\tz$ ($\omtau t = 100$). Shortly after that, starting at $\omtau t=105$, the orange circles abruptly deviate from the black line. Analogous comparisons between the eNRG and tight-binding results for the plugged components ($\ffp$) and unplugged components ($\ffu$) of the fidelity (not shown) ratify this finding and show that even with the relatively large $\lambda=2$ and without averaging over $\theta$ or $\zeta$, the eNRG procedure yields quantitatively accurate results for $t < \tz$.

  \begin{figure*}
  \centering
  \includegraphics[width=1.6\columnwidth]{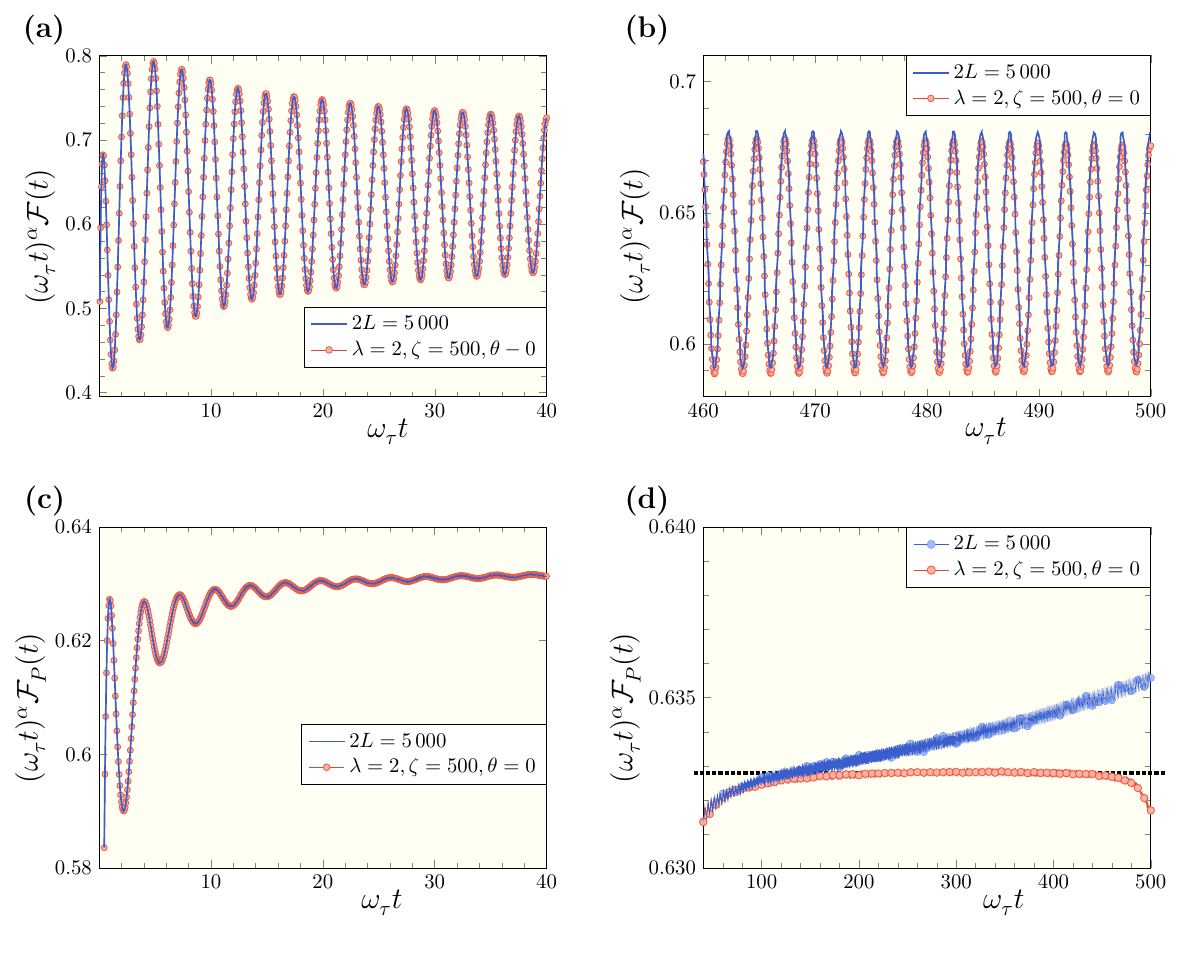}
  \vspace{-0.75cm}
  \caption[4fidelities]{Comparison between the fidelities for the eNRG and tight-binding Hamiltonians with $K=-2\tau$. The inset boxes display the length of the tight-binding lattice and the parameters defining the eNRG Hamiltonian truncated at $\nn =21$. The tight-binding results came from the diagonalization that yielded the data in Figs.~\ref{fig:6}~and \ref{fig:7}. Panels (a) and (b) compare the fidelities at short and long times, respectively, to contrast agreement in the former range with small disagreement in the latter. Panels (c) and (d) display the equivalent comparisons for the plugged contributions to the fidelities. Since the product $(\omega_\tau t)^{\alpha}\ffp(t)$ must approach a constant at long times, panel (d) proves the $\zeta=500$ eNRG results more precise than the $ 2L=5\,000$ tight-binding data.}
  \label{fig:8}
\end{figure*}

Figure~\ref{fig:8} extends the comparison to longer times $0 \le \omtau t < T_{\mbox{max}}\z = 500$. To check the exactness of the procedure with the smallest offset compatible with the requirement $\tz\ge T_{\mbox{max}}\z$, we have computed the fidelities with $\zeta=500$. Panel (a) shows excellent agreement between the eNRG results at short times ($\omtau t < 50$) and those derived from the diagonalization of the tight-binding Hamiltonian with $ 2L=5\,000$. At longer times ($\omtau t \alt 500$), however, the agreement deteriorates, as panel (b) shows. To identify the origin of the discrepancy, panels (c) and (d) show the contributions of the plugged many-body states to the fidelities divided by the Doniach-Sunjic power law for both Hamiltonians over different ranges of time. In analogy to panel (a), panel (c) witnesses the excellent agreement between the plugged fidelities at short times ($\omtau t < 40$). Panel (d) expands the vertical axis to zoom in near the $t\to\infty$ limit, indicated by the dashed black line. Starting at $\omtau t=180$, the filled blue circles that represent the results for the tight-binding Hamiltonian rise above the black line. At $t=500 = T_L/10$, the blue curve is 0.6\% above the continuum limit. The filled orange circles, which represent the plugged fidelity computed with $\zeta=500$, agree with the blue circles at short times and follow the black line until $\omtau t=450$ when deviations become visible. At $t=\tz=500$, the deviation is 0.2\%. In summary, the disagreement between the blue and black curves exposes a finite-size effect that pushes the fidelities $\ff(t)$, $\ffp(t)$, and $\ffu(t)$ (not shown) for the tight-binding Hamiltonian away from the continuum limit for $\omtau t \agt T_L\z/20$. The eNRG results computed with large $\zeta$ are in excellent agreement with the continuum up to $t\approx 0.95\,\tz$ and reach the maximum relative deviation of 0.2\% at $t=\tz$. In contrast, small-size effects arise in eNRG calculations with small $\zeta$, as discussed in the following section.

\subsection{Smoothed fidelity procedure}
\label{sec:average-fidelity}

As Sec.~\ref{sec:large-zeta} explained, the eNRG procedure yields essentially exact fidelities in the time range $0\le t< \tz$. From the perspective of computational cost, the eNRG method is very efficient. Given a time interval, the effort needed to compute the fidelity is a small fraction of the effort to diagonalize the tight-binding Hamiltonian that yields comparable accuracy. For example, the eNRG run and the diagonalization of the $ 2L=5\,000$ tight-binding Hamiltonian that provided the data in Fig.~\ref{fig:8} demanded 42 seconds and 7100 seconds of CPU time, respectively, on a standard laptop computer.

These considerations refer to our model Hamiltonian, which reduces to two quadratic forms. The diagonalization of a quartic eNRG Hamiltonian would be substantially more demanding computationally. Large $\zeta$'s would make the cost of diagonalizing a Hamiltonian of the latter kind, iteratively~\cite{Wilson_NRG}, prohibitive because the dimension of the basis that defines the eNRG Hamiltonian~\eqref{eq:73} is $\zeta +\nn+1$, and the spectrum of the eNRG Hamiltonian cannot be truncated in the first $\zeta$ iterations. In practice, therefore, we are restricted to moderate offsets, that is, $\zeta$'s that are substantially smaller than the time interval of interest. This limitation brings us to the problem of computing $\ff(t)$ for $t>\tz$.

In the time domain defined by this inequality, we have to deal with beat frequencies like those in Fig.~\ref{fig:6}. To this end, following the procedure proven reliable in the computation of thermodynamic and transport properties \cite{2022FeO075129}, we average the raw fidelities on a set $C_{\zeta}\z$ containing $N_\zeta\z=2$ consecutive offsets $\zeta$ and integrate them over $\theta$ to obtain the \emph{smoothed} fidelity:
\begin{align}
  \label{eq:82}
  \bar{\ff}(t) = \dfrac{1}{N_{\zeta}\z}\sum_{\zeta\in C_{\zeta}\z}\int_{-1}^1\fz (t) \,\dd\theta.
\end{align}

Here, we work with $C_{\zeta} = \{3,4\}$ and use Simpson's $\sfrac13$ rule with an even number $N_\theta\z$ of intervals to evaluate the integral on the right-hand side of Eq.~\eqref{eq:82}, that is, a set of $N_\theta\z +1$ equally spaced points ranging from $\theta=-1$ to $\theta=1$.

\begin{figure*}[!ht]
  \centering
  \includegraphics[width=1.6\columnwidth]{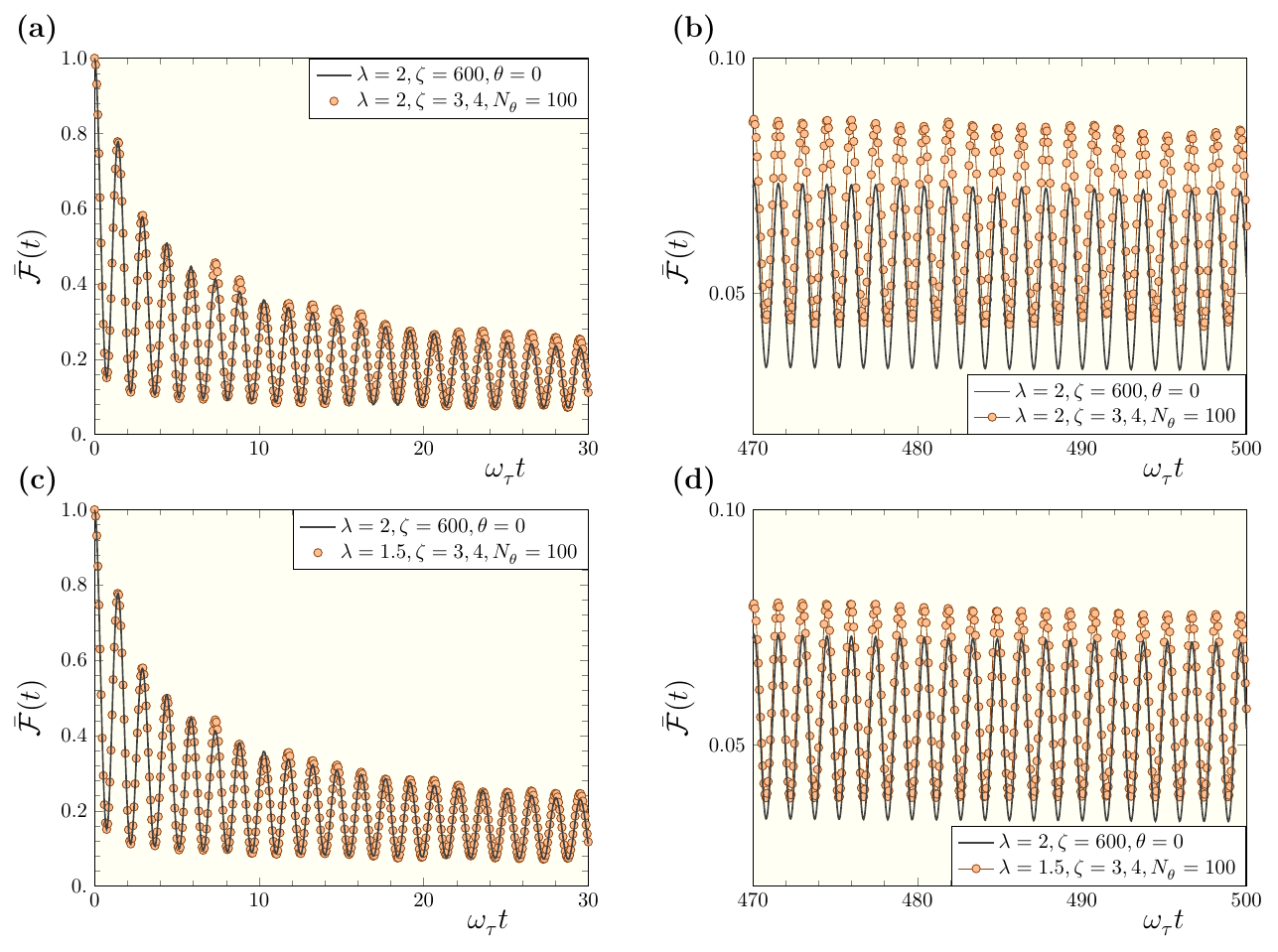}
  \vspace{-0.25cm}
  \caption[4fid]{\label{fig:9}Comparison between time-dependent fidelities for the potential scattering $K=-4\tau$ computed with distinct sets of eNRG parameters. In each panel, the black curve
  represents essentially exact data for the fidelity, computed with $\lambda=2$, $\zeta = 600$, $\theta=0$, and $\nn=21$, and the filled orange circles represent the smoothed fidelity $\overline{\ff}_{\zeta,\theta}$, averaged on $\zeta=3,4$ and integrated over $\theta$ with $N_\theta\z= 100$. Panels (a) and (b) show the short- and long-time behaviors, respectively, computed with $\lambda=2$ and $\nn=22$. Panels (c) and (d) show the equivalent results computed with $\lambda=1.5$ and $\nn=27$.}
\end{figure*}

Figure~\ref{fig:9} shows the time-dependent fidelity for $K=-4\tau$, calculated in runs with $N_\theta=100$ and two discretization parameters: $\lambda=2$ [panels (a) and (b)] and $\lambda=1.5$ [panels (c) and (d)]. Each panel compares orange circles representing the smoothed fidelities with a solid black curve that shows the essentially exact fidelity computed by the procedure in Sec.~\ref{sec:large-zeta} with $\zeta=600$.

At small times [panels (a) and (c)], the circles match the black curve fairly well. At larger times, in panels (b) and (d), substantial deviations arise. Although each curve in these panels oscillates around a constant, the averages and amplitudes of the oscillations drawn by the orange circles are visibly larger than those of the solid black curves.

\begin{figure}[!ht]
  \centering
\includegraphics[width=0.85\columnwidth]{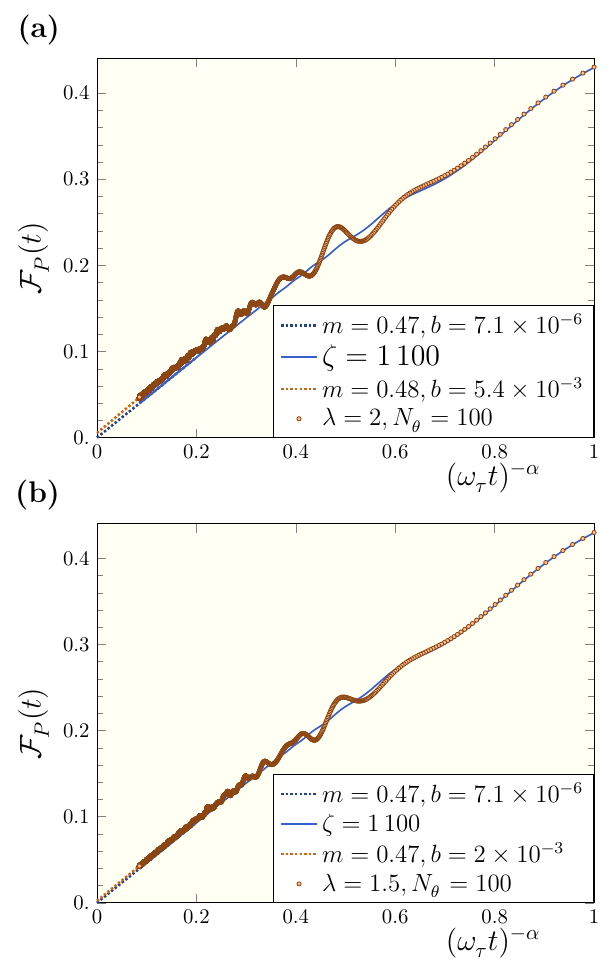}
  \caption[long-timeF]{\label{fig:10} Comparison between the long-time behaviors of the plugged fidelity for $K=-4\tau$ computed with different sets of eNRG parameters and plotted as functions of the Doniach-Sunjic power law $(\omega t)^{-\alpha}$. In the two panels, the solid blue curve represents essentially exact results for the fidelity, computed with $\lambda=2$, $\zeta=1\,100$, $\theta=0$ and $\nn = 23$, while the dotted blue line shows its extrapolation to $t\to\infty$. The solid orange circles in panel~(a)~[(b)] represent the plugged fidelity computed with $\lambda=2$ ($1.5$), averaged on $\zeta=3,4$ and integrated with $N_{\theta}\z=100$. The dotted orange lines depict the linear extrapolation of the orange curves to $t\to\infty$. The legends show the linear intercept $m$ and slope $b$ of each extrapolation.}
\end{figure}

At the root of these discrepancies are residual contributions analogous to the linear intercept in the inset of Fig.~\ref{fig:5}. It follows from Eq.~\eqref{eq:18} that numerically computed fidelities approach the constant $\cres$, defined by \eqref{eq:19}, as $t\to\infty$. Likewise, the contributions $\ffp(t)$ [$\ffu(t)$] of the plugged states (unplugged) to the fidelity must approach the corresponding residual values $\cresp$ ($\cresu$). To confirm these expectations, Fig.~\ref{fig:10} shows $\ffp$  as a function of the Doniach-Sunjic function $(\omega_\tau t)^{-\alpha}$ and the linear extrapolations of $\ffp$ to $t\to\infty$ for $\lambda=2$ [panel~(a)] and $\lambda=1.5$ [panel~(b)]. As in Fig.~\ref{fig:5}, the dotted lines representing the linear regressions fit the computed fidelities very well for $\omtau t>100$, which corresponds to $(\omtau t)^{-\alpha}\approx 0.19$, and the intercepts determine the residual constants, $\cresp = \num{5.4e-3}$ for $\lambda=2$, and $\cresp=\num{2.2e-3}$ for $\lambda=1.5$. For comparison, the two panels also show the fidelity calculated with $\zeta=1\,100$ and its extrapolation to $t\to\infty$, which yields a much smaller residual constant,  $\cresp = \num{7.1e-6}$. The plots of $\ffu(t)$ as a function of the Nozière-De Dominicis power $(\omega_\tau t)^{-\beta}$ (not shown) yield analogous results.

In summary, the calculated fidelities oscillate about a residual constant $\cres$ at large times. As $t\to\infty$, the progressive damping of the oscillations brings $\bar{\ff}(t)$ to this residual value. Whereas large-offset computations yield very small $\cres$, the larger residual constants resulting from small-offset, large $N_{\theta\z}$ runs introduce significant vertical displacements in plots such as those in Fig.~\ref{fig:9}(b) and Fig.~\ref{fig:9}(d).

The magnitudes of the discrepancies depend on $\lambda$. The differences in panel~(d) are substantially smaller than those in panel~(b), a reduction indicating that the computed fidelities rapidly converge to the tight-binding ($\lambda=1$) limit. To monitor this convergence, Secs.~\ref{sec:dependence-n_thetaz}~and \ref{sec:dependence-lambda} study the dependence of $\bar{\ff}(t)$ on the eNRG parameters. For easier comparison, instead of the computed fidelities $\bar{\ff}(t)$, we will examine the deviations
\begin{align}
  \label{eq:84}
  \Delta \ff(t) \equiv \bar{\ff}(t) - \fr(t),
\end{align}
where the reference function $\fr(t)$ is the fidelity for $K=-4\tau$ calculated with $\lambda=2$, $\zeta=1\,100$, and $\theta=0$, a parametric choice that yields essentially exact results in the interval $0\le \omtau t<1\,000$.

\subsubsection{Dependence on $N_{\theta}\z$}
\label{sec:dependence-n_thetaz}

\begin{figure}[!ht]
  \centering
  \includegraphics[width=0.8\columnwidth]{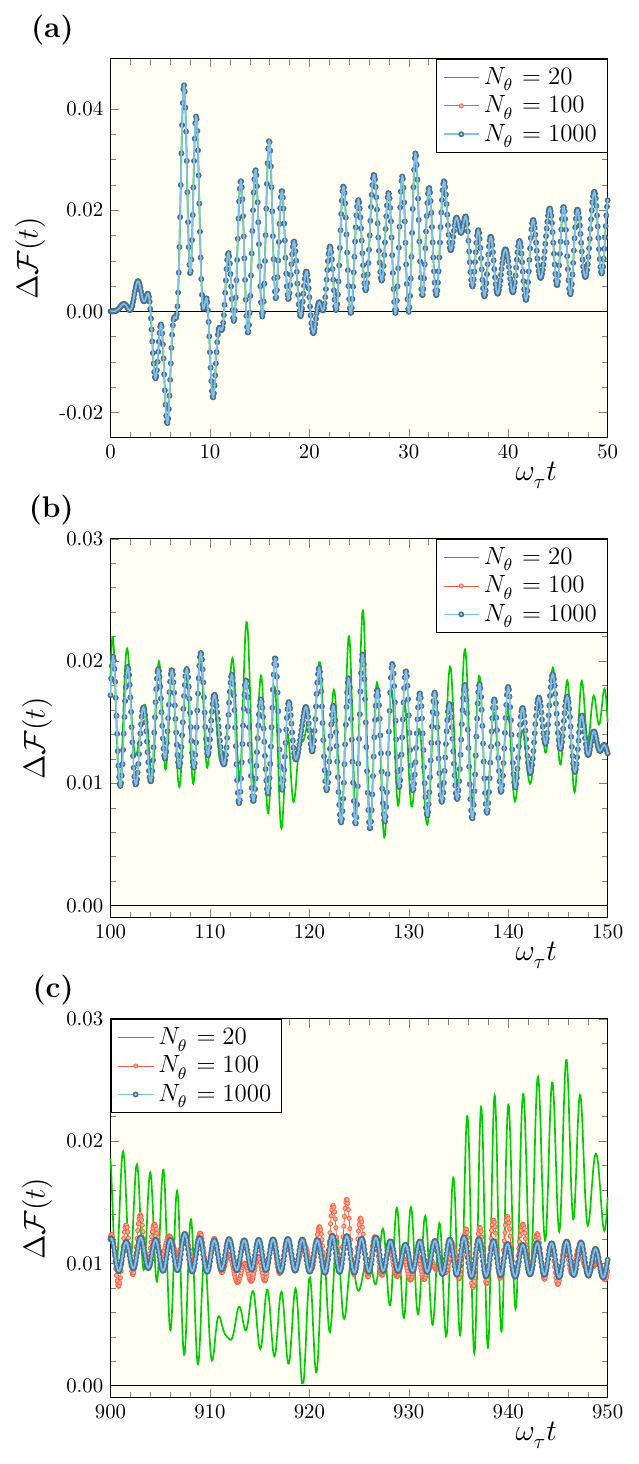}
  \caption[3thetas]{Deviations of the fidelities computed for $K=-4\tau$ with $\lambda=2$, averaged on $\zeta\in \{3,4\}$ and the indicated numbers of $\theta$'s ($N_{\theta}\z=20$, solid green line; $N_{\theta}\z=100$, orange circles; and $N_{\theta}\z=1\,000$, blue circles) from the essentially exact curve, computed with $\lambda=2$, $\zeta=1\,100$, and $\theta=0$, at short [panel (a)], medium [panel (b)], and long times [panel (c)]. }
  \label{fig:11}
\end{figure}

In the $L\to\infty$ limit, the spectrum of the tight-binding Hamiltonian constitutes a continuum. By contrast, the discrete single-particle eigenvalues~\eqref{eq:74} of the eNRG Hamiltonian introduce artificial splittings; as a result, $\fz(t)$ oscillates unphysically. The integration in $\theta$ on the right-hand side of Eq.~\eqref{eq:82} aims to smooth out these spurious deviations. In practice, however, the integral is evaluated numerically via Simpson's rule, and the result depends on the number of subdivisions $N_{\theta}\z$, or the difference $\Delta\theta = \sfrac{2}{N_\theta\z}$ separating successive integration points. If $N_\theta\z$ is sufficiently large, so that the energy splitting $\Delta E$ between two eigenvalues~\eqref{eq:74} calculated at $\theta$ and $\theta+\Delta\theta$ is small compared to the energy resolution $\sfrac{\hbar}t$ set by the uncertainty principle, Simpson's rule should be accurate and produce smoothed fidelities $\ff(t)$ independent of $N_\theta\z$. At longer times, the resolution will be finer and $N_\theta\z$ may have to be increased. 

Figure~\ref{fig:11} confirms this reasoning. The solid green lines and the orange and blue circles show the deviations $\Delta \ff(t)$ resulting from eNRG calculations of $\bar{\ff}(t)$ for $K=-4\tau$ with $N_{\theta}\z =20, 100$ and $1\,000$, respectively. The three runs were carried out with $\lambda=2$ and averaged on $\zeta\in\{3,4\}$. Panel (a) covers the interval $0\le \omtau t <50$ and shows three congruent curves, demonstrating that in this interval the energy uncertainty $\sfrac{\hbar}{t}$ dwarfs the energy differences $\Delta E$. Even the largest differences, associated with $N_{\theta}\z=20$, are much smaller than $\sfrac{\hbar}{t_{\mbox{max}}\z}$, ($\omtau t_{\mbox{max}}\z=50$). In the intermediate time interval depicted in panel (b), the broadening becomes smaller than the energy splittings $\Delta E\z$ for $N_{\theta}\z=20$; the discrepancies between the green line and the circles reflect the deviations caused by Simpson's approximation.

Panel (c) depicts the interval $900<\omtau t<950$. Now, the energy uncertainties are smaller than the $\Delta E$ for $N_{\theta}\z=20$ and $100$, but larger than the splittings for $N_{\theta}= 1\,000$. Accordingly, the blue circles draw a nearly uniform, small-amplitude oscillation, in contrast with the less regular behavior of the red circles and the prominent swings of the green line.

Altogether, the three panels indicate that with sufficiently large $N_{\theta}\z$---as a rule of thumb, $N_{\theta}\z$ must exceed $\omtau t$---integration over $\theta$ eliminates the spurious oscillations due to the discrete the eNRG spectra; the minute, regular deviations depicted by the blue circles in panel (c) offer especially compelling evidence. 

However, the fluctuations in panel (a) defy smoothing, with alternating positive and negative values that can be as large as \num{5e-2} at short times. The deviations become systematically positive and more harmonic as time increases. These oscillations, drawn by all three curves in panel (a), by the curves calculated with $N_\theta\z=100$ and $1\,000$ in panel (b), and by the curve with $N_\theta\z=1\,000$ in panel (c), cannot be attributed to the nonuniformity of the single-particle spectrum.

Therefore, Fig.~\ref{fig:11}  identifies two classes of deviations. In the first class are those due to nonuniformity, which persist indefinitely and will hence be called \emph{persistent}. The fluctuations of the orange curve in the inset of Fig.~\ref{fig:6} constitute an example. Simpson integration over $\theta$ with sufficiently large $N_\theta\z$ combined with averaging over $\zeta$ smooths out the discrepancies in this class.  

The second class comprises deviations that decay but cannot be smoothed out, as best illustrated by the blue circles in Fig.~\ref{fig:11}. We call them \emph{transitory}. To identify the source of these discrepancies, Sec.~\ref{sec:dependence-lambda} compares eNRG data for different discretization parameters. 

\subsubsection{Dependence on $\lambda$}
\label{sec:dependence-lambda}
As explained in Sec.~\ref{sec:time-scales}, each of the two functions $\bar{\ff}(t)$ and $\fr(t)$ on the right-hand side of Eq.~\eqref{eq:84} receives a contribution $\ffp(t)$ from the plugged states and a contribution $\ff_U\z(t)$ from the unplugged ones. Since $\ff_P\z(t)$ and $\ffu(t)$ behave differently and approach different residual constants at long times, the function $\Delta\ff(t)$ depends on all the eNRG parameters.

The most important for our purposes is the dependence on the discretization parameter $\lambda$. As mentioned in Sec.~\ref{sec:single}, the discretization leading to Eq.~\eqref{eq:80} is an approximation that can only be justified \emph{a posteriori}. To this end, we must show that the eNRG computation of the physical property of interest converges rapidly to the continuum limit ($\lambda=1$) so that calculations carried out with discretization parameters compatible with practical considerations of computational cost yield absolute deviations of a few percent or less.
% In the context of the photoemission problem, 3\% absolute deviations $\abs{\Delta\ff(t)}$ are tolerable in computations with $\lambda\ge \sqrt{2}$ throughout the time interval $0< \omtau t <1\,000$.

\begin{figure}[!ht]
  \centering
\includegraphics[width=0.95\columnwidth]{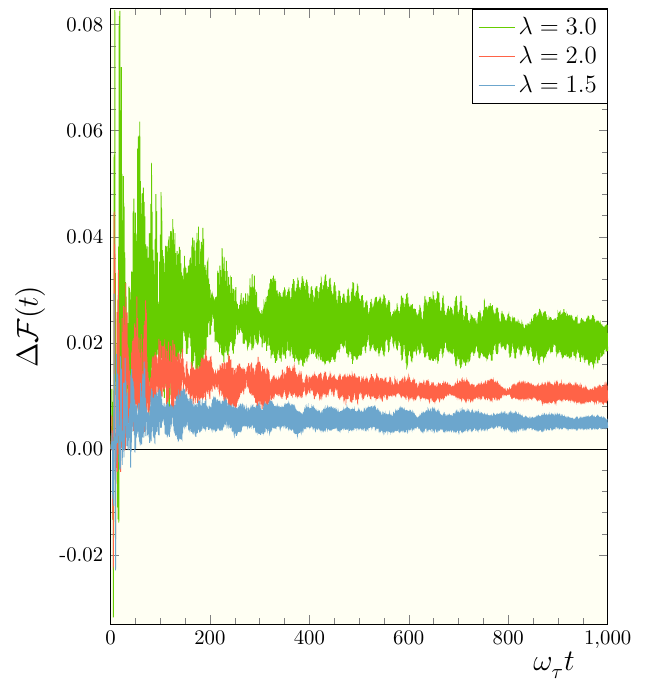}
  \caption[3lambdas]{Deviations of the fidelities computed for $K=-4\tau$ with $\zeta \in \{3,4\}, N_\theta=1\,000$, and the indicated discretization parameters from the essentially exact results obtained with $\lambda=2$, $\zeta=1\,100$, $\theta=0$, and $\nn = 23$.}
  \label{fig:12}
\end{figure}

Figure~\ref{fig:12} shows the deviations resulting from three eNRG runs for $K=-4\tau$ with distinct discretization parameters: $\lambda = 1.5$ ($\nn=38$, solid blue curve), $2$ ($\nn = 22$, red), and $3$ ($\nn =14 $, green). All fidelities were averaged on $\zeta\in\{3,4\}$ and integrated with $N_{\theta}\z =1\,000$. Panels~(a), (b),~and (c) in Fig.~\ref{fig:11} can be regarded as blow-ups of the red curve in Fig.~\ref{fig:12} that expand the intervals $0\le\omtau t\le 50$, $100\le\omtau t\le 150$, and $900\le\omtau t\le 950$, respectively.

In the interval $0\le\omtau t\le10$, the three curves exhibit relatively large variations. The green curve fluctuates between $\Delta \ff =-\num{3e-2}$ and $\num{8e-2}$, approximately, the red curve, between $\Delta \ff =-\num{2e-2}$ and $\num{4e-2}$ [cf.~Fig.~\ref{fig:11}(a)], and the blue curve, between $\Delta \ff=-\num{2e-2}$ and $\num{3e-2}$. Fluctuations become more regular as $\omtau t$ grows and their amplitudes decrease. At the end of the interval, $\omtau t\approx1\,000$, each curve is centered on a positive constant, approximately equal to $\num{2e-2}$ for $\lambda=3$, $\num{1e-2}$ for $\lambda=2$, and $\num{5e-3}$ for $\lambda=1.5$. The amplitudes also decay as $\lambda$ decreases, at all times. Thus, the eNRG procedure with $\lambda=1.5$ produces accurate smoothed fidelities, except for the short time interval $0\le \omtau t <200$. Outside that range, the absolute deviations never reach $\num{1e-2}$.

In the light of renormalization-group theory \cite{1980KWW1003,1980KWW1044,RevModPhys.55.583,Wilson1974}, Fig.~\ref{fig:12} helps to identify the source of transitory fluctuations. The substantial deviations resulting from the run with $\lambda=3$ point to operators artificially introduced by the eNRG discretization. Given the Hamiltonian $\hh_K^\nn$, the mapping $\mathbb{T}[\hh_K^\nn]= \hh_K^{\nn+2}$ defines the renormalization-group transformation $\mathbb{T}$ \cite{1980KWW1003}. As $\nn$ increases, $\hh_K^\nn$ approaches a fixed point $\hh_{\delta}^*$ of the transformation, characterized by the phase shift in Eq.~\eqref{eq:25} \cite{1980KWW1044}. At given $\nn$, the difference $\delta_\hh^\nn = \hh_K^\nn - \hh_{\delta}^*$ can be written as a linear combination of charge-conserving operators in the space spanned by the many-body eigenstates of $\hh_K^\nn$:
\begin{align}
\label{eq:85}
\delta_\hh^\nn = \sum_{m=0}^\mathcal{M} w_m^\nn\mathcal{O}_m\z,
\end{align}
where $\mathcal{M}$ is the number of all such operators. 

Equation~\eqref{eq:85} is in diagonal form. The $\mathcal{O}_m\z$ on the right-hand side are eigenoperators of the transformation $\mathbb{T}$, termed \emph{irrelevant} because the corresponding eigenvalues are smaller than unity. As $\nn\to\infty$, the coefficients $w_m^\nn \to0$, and the right-hand side of Eq.~\eqref{eq:85} vanishes. 

The eigenvalues and eigenvectors of the fixed-point Hamiltonian depend on the discretization parameters $\lambda$, $\zeta$, and $\theta$, and so do the physical properties derived from them. The time dependence of the raw fixed-point fidelity shows artificial oscillations, analogous to those in Fig.~\ref{fig:6}, which disappear after $\ff_{\zeta,\theta}\z(t)$ is smoothed over $\zeta$ and $\theta$. More importantly in the context of the present discussion, the residual constant defined by Eq.~\eqref{eq:19}, which vanishes in the continuum, increases monotonically with $\lambda$. The linear regressions in Fig.~\ref{fig:10} offer illustrations.

Figure~\ref{fig:12} offers additional evidence. At long times ($\omtau t>800$), each curve fluctuates about the residual constant $\cres$ associated with the corresponding discretization parameter $\lambda$. The blue, orange and green curves, calculated with $\lambda=1.5$, $2$ and $3$, fluctuate about $\cres= \num{5e-3}$, $\num{1e-2}$ and $\num{2e-2}$, respectively.

The irrelevant operators in $\hh_K^\nn$ account for the large fluctuations at smaller times. The decay of the irrelevant operators with $\nn$ defines a hierarchy based on their eigenvalues. The leading irrelevant eigenoperator of this quadratic Hamiltonian is \cite{1980KWW1003}
\begin{align}\label{eq:86}
\mathcal{O}_1\z\equiv f_0^\dagger f_1+ \hc,
\end{align}
with eigenvalue $\lambda^{-2}$, that is,
\begin{align}\label{eq:87}
\mathbb{T}[\mathcal{O}_1\z] = \lambda^{-2}\mathcal{O}_1\z;
\end{align}
therefore, the coefficient $w_1^\nn$ follows the decay of the energy. 

In analogy to $\cres$, the coefficients $w_m^\nn$ ($m=1,\ldots,\mathcal{M}$) of the irrelevant operators increase in magnitude with $\lambda$ \cite{1980KWW1003}. Therefore, these operators distinguish $\hh_K^\nn(\lambda)$ from the tight-binding Hamiltonian and push the calculated physical properties away from their continuum values. Consequently, the green curve in Fig.~\ref{fig:12} ($\lambda=3$) shows deviations larger than the orange curve ($\lambda=2$), and the latter, larger than the blue curve ($\lambda=1.5$).

To explain the relatively slow time decay of the fluctuations in all three curves, a comparison with equilibrium properties is expedient. The calculation of equilibrium properties deals with a single energy scale at a time: thermodynamic and equilibrium transport properties are associated with the thermal energy scale $k_B\z T$; spectroscopy, with the photon energy $\hbar\omega$. This energy scale defines the truncation number $\nn$, in analogy with Eq.~\eqref{eq:72}, with the pertinent inverse energy, $\sfrac{1}{k_B\z T}$ or $\sfrac{1}{\hbar \omega}$, substituted for $\sfrac{t}{\hbar}$. The deviation of a physical property caused by the leading operator $\mathcal{O}_1\z$ on a given energy scale is then proportional to its coefficient $w_1^\nn$ and therefore decays with energy. The deviations are, therefore, insignificant, except at energies comparable to the bandwidth.

Time-dependent properties pose a different problem. The uncertainty principle broadens the energy levels by $\Delta_t\z = \sfrac{\hbar}{t}$. Instead of an energy scale, this broadening defines a soft cutoff \cite{2005AnS196801}: the energy scales above $\Delta_t\z$ contribute to $\bar{\ff}(t)$, although their contribution is a gradually decreasing function of $t$. At very short times, the sum of all irrelevant operators gives rise to substantial deviations $\Delta \ff$. As time increases, the truncation number $\nn$ increases, the coefficients of the irrelevant operators decrease, and the operator $\mathcal{O}_1\z$ becomes dominant. However, the uncertainty principle admits contributions from higher energy scales, which correspond to smaller truncation numbers $\nn$ and hence include other irrelevant operators; as a result, $\Delta \ff$ decays more slowly than $\sfrac{1}{\omtau t}$, as illustrated by the plots in Fig.~\ref{fig:12}.

\subsection{Retrospective overview}
\label{sec:retro-overview}
The two real numbers $\lambda$ and $\theta$ and the integer $\zeta$ define the eNRG Hamiltonian. The real numbers are analogous to the discretization and twist parameters of the NRG method $\Lambda$ and $z$, respectively. The integer $\zeta$, characteristic of the eNRG approach, defines two alternative ways to calculate time-dependent properties, which we briefly recapitulate. For definiteness, let $\tmax$ specify the time interval $0\le t\le \tmax$ of interest.

\subsubsection{Large $\zeta$ procedure}
\label{sec:large-zeta_}
Choosing $\zeta$ somewhat larger than $\omtau \tmax$ with moderately small $\lambda$ ($\lambda=2$, for example) and $\theta=0$ warrants uniform precision over the entire time interval. A single run being sufficient, this procedure is recommended for quadratic Hamiltonians, which can be easily diagonalized. For more complex models of correlated impurity systems, which require iterative diagonalization of the discretized Hamiltonian, the computational cost tends to be prohibitive. If $\tmax$ is only moderately large, the tDMRG method applied for the eNRG Hamiltonian offers an attractive alternative.

\subsubsection{Integration over twist parameters}
\label{sec:n_theta}
When the computational cost considerations rule out double-digit $\zeta$'s, it becomes convenient to carry out several runs, with $N_\theta\z$ uniformly spaced twist parameters in the interval $-1\le\theta\le1$. To minimize discretization artifacts, it is expedient to average the computed time-dependent property on two subsequent $\zeta$'s and to integrate it numerically over $\theta$. The calculated properties tend to show artificial oscillations around $\omtau t\gtrsim\zeta$, which decay with time, resulting in accurate results for $\omtau t\lesssim N_{\theta}\z$.

\section{\label{sec:conclude}Conclusions}
We have studied the time dependence of the fidelity $\ff(t)$, the Fourier transform of X-ray photoemission spectra. The two general approaches in Secs.~\ref{sec:many-body-procedure}~and \ref{sec:single-part-meth} determine  $\ff(t)$. The former has led to the analytical expression derived in Appendix~\ref{sec:appendix25}, which is very accurate for $\omtau t\gg 1$. The latter, which is convenient for numerical computations, has allowed the calculation of $\ff(t)$ from the eigenvalues and eigenvectors of the tight-binding Hamiltonian for the photoemission problem. 

The analytical expression and numerical results of this calculation have served as benchmarks against which we have gauged the precision of the two time-dependent renormalization-group procedures defined in Sec.~\ref{sec:enrg}. The results and comparison offer physical and operational insight.

Our findings highlight the importance of the bound state created by the core-hole potential. As a function of time, the fidelity decays following a power law enveloped by a damped oscillation. The decay is the Doniach-Sunjic power law. The oscillations are due to interference between two kinds of excitation: excitations to plugged states, with the bound state occupied, and to unplugged states, with the bound state vacant. The oscillations are damped because the latter follows the Nozières-De Dominicis power law and, therefore, decays faster than the former, which follows the Doniach-Sunjic law.

We have rooted our developments in the real-space formulation of the renormalization-group approach, the eNRG method, which generates model Hamiltonians parametrized by two real numbers $\lambda$ and $\theta$, and an integer $\zeta$. Comparison with benchmarks identified two sources of inaccuracy affecting the fidelities calculated by the eNRG procedures, both due to the discretization of the conduction band. The first source is the irrelevant operators that distinguish the eNRG model Hamiltonians from the fixed points to which the renormalization-group transformation drives them. The irrelevant operators cause slowly decaying transitory deviations. The second source is the reduction of the band continuum to a sequence of discrete levels, which spawns persistent fluctuations and shifts upward the calculated fidelities.

The large-$\zeta$ procedure, which amounts to working with $\zeta> \omtau t_{\mbox{max}}\z$, where $t_{\mbox{max}}\z$ is the longest time of interest, produces essentially exact results over the interval $[0,t_{\mbox{max}}\z]$,  and therefore avoids deviations.

The large-$\zeta$ algorithm deals so efficiently with quadratic Hamiltonians and yields so precise fidelities that it has replaced the brute-force procedure and established a more reliable standard of accuracy in our study. We have checked the quality of the fidelities produced by the second eNRG method, the smoothing procedure, against this new benchmark. The second algorithm is valuable because it offers better perspectives for work targeting Hamiltonians with quartic terms. The inability to eliminate transitory deviations limits its range to medium and large times, to wit, the region that may be beyond the scope of the large-$\zeta$ method. 

{The significant contribution of the high-energy states, referred to here as unplugged states, arises from the sudden change in a localized interaction and the presence of a bound state. These features are not exclusive to quadratic Hamiltonians and can also play an important role in interacting systems \cite{cornaglia2007}. Discarding states based on energy is therefore inadequate in such cases. More sophisticated truncation schemes than the standard ultraviolet cutoff are required to compute their properties within the many-body iterative diagonalization procedure \cite{2005AnS196801}. Combined with an appropriate truncation strategy, the second algorithm may provide a route to accurately access long-time dynamics in impurity models. Further developments are left for future work.}

In summary, we studied the time dependence of the photoemission current as an object of physical interest in its own right and as a testing ground for the development of time-dependent procedures based on the eNRG method. The analytical expression for the fidelity and the large-$\zeta$ algorithm offered (i) solutions to the problem, with results that are uniformly accurate over the one-dimensional parametric space of the model and (ii) a physical interpretation of all features of the solution. With future applications to interacting-impurity Hamiltonians in mind, we have also discussed the smoothing algorithm. Our results show that averaging on $\zeta$ followed by integration over $\theta$ eliminates persistent deviations and is therefore appropriate to calculate time-dependent properties after the transient deviations have died out.

\lno{\textbf{Data Availability Statement}
The data that support the findings of this article are openly available \cite{2025DinLSB9}.
}

\acknowledgments{ - The authors thank Hong-Hao Tu and Jan von Delft for suggesting the single-particle method outlined in Sec.~\ref{sec:single-part-meth}. This research received financial support from FAPESP (grant 2022/05198-2), the CNPq (grant 311689/2023-0), and the Coordenação de Aperfeiçoamento de Pessoal de Nível Superior – Brasil (CAPES) – Finance Code 001. FDP's work was supported by PhD and internship fellowships from FAPESP ( grants 2022/09312-4 and 2024/05637-1), GD's by a PhD scholarship from the Brazilian agency Coordenação de Aperfeiçoamento de Pessoal de Nível Superior (CAPES), and MPL's by an undergraduate research fellowship from FAPESP (2021/11154-5). I.D. thanks the Instituto de Fısica de S\~ao Carlos, University of S\~ao Paulo, Brazil, for the kind hospitality. We gratefully acknowledge the support and access to computational resources offered by the Center for Mathematical Sciences Applied to Industry (CeMEAI), funded by FAPESP (grant 2013/07375-0).}

\appendix

\section{Derivation of Eq.~\eqref{eq:56}}
\label{sec:appendix25}
The derivation of the analytical expression~(\ref{eq:56}) for the \textit{fidelity}, $\ff(t)$, involves several approximations that become reliable under the conditions $\omtau t\gg 1$ and $L\gg 1$. The first approximation amounts to replacing the tight-binding dispersion relation~(\ref{eq:22}) by the linear form~(\ref{eq:23}), which we rewrite here in the equivalent form
\begin{align}\label{eq:90_}
  \varepsilon_q\z = \dl\left(q +\dfrac12\right) \qquad \left(q = -L, -(L-1), \ldots, L-1\right),
\end{align}
\noindent with $\Delta_L\equiv\tau\pi/L$. This approximation loses accuracy near the borders of the conduction band, that is, for $q\approx \pm L$, a shortcoming that affects the precision of Eq.~(\ref{eq:56}) at short times only.

For consistency, we also adopt an expression equivalent to the linearized Eqs.~(\ref{eq:25}), which represent the single-particle eigenvalues of the final Hamiltonian:
\begin{align}
  \label{eq:91_}
    \epsilon_{\ell}\z &= \dl \left(\ell+\dfrac12- \dfrac{\delta}{\pi}\right) \nonumber \\ \Big(\ell=-(L-&1),-(L-2),\ldots, L-1\Big), 
\end{align}
and
\begin{align}
\label{eq:910_}
\epsilon_{-L}\z = 
\begin{cases}
    \dl(-L+\dfrac{1}{2}-\dfrac{\delta}{\pi})&\qquad(\abs{K}<\tau)\\[2pt]
    -(K + \dfrac{\tau^2}{K})&\qquad(\abs{K}\ge \tau).
\end{cases}
\end{align}

In addition, we let the phase shift $\delta$ be the constant given by Eq.~(\ref{eq:26}), although that equality is only exact at the Fermi level.

Under these assumptions, the commutator between an initial eigenstate $a_{q}\dg$ and a final one $b_{\ell}\z$ is given by Eq.~(\ref{eq:29}), to wit,
\begin{align}
  \label{eq:92_}
  \acomm{a_q\dg}{b_{\ell}\z} =\dfrac{\sin(\delta)}{\pi} \dfrac{1}{q-\ell  +\dfrac{\delta}{\pi}},
\end{align}
for all $q$ and $\ell$, except $\ell=-L$.

As explained in Sec.~\ref{sec:numerical-results}, the many-body eigenstates of the final Hamiltonian can be divided into two disjoint classes of plugged $\ket{\rp}$ and unplugged $\ket{\ru}$ states. In this notation, Eq.~(\ref{eq:18}) reads
\begin{align}
  \label{eq:93}
  %\mathcal{F}(t)=\left|\braket{\Psi_0\z}{\Psi(t)}\right|^2 =
  \mathcal{F}(t)=&\Bigg|
  \sum_{r_P\z}\z\abs{\braket{r_P\z}{\Psi_0\z}}^2 \exp(-i\dfrac{E_{r_P\z}}{\hbar}t) 
   \nonumber \\ 
  &+\sum_{r_U\z}\z\abs{\braket{r_U\z}{\Psi_0\z}}^2 \exp(-i\dfrac{E_{r_U\z}}{\hbar}t)\Bigg|^2,
\end{align}
\noindent where $\ket{\Psi_0}$ is the many-body ground state of the initial Hamiltonian ($K=0$).

\subsection{Plugged eigenstates}
\label{sec:many-body-eigenst}
In the many-body ground state of the initial Hamiltonian, all single-particle levels $\varepsilon_n\z<0$, below the Fermi level ($n=-1,-2,\ldots,-L$), are occupied, while the other levels remain vacant. Therefore,
\begin{align}
  \label{eq:94}
  \ket{\Psi_0\z} = \prod_{j=-L}^{-1} a_{j}\dg\ket{\emptyset}. 
\end{align}
where $\ket{\emptyset}$ is the vacuum state.

Likewise, in the many-body ground state $\ket{\Omega}$ of the final Hamiltonian, all single-particle levels $\epsilon_n\z$  below the Fermi level ($n=-1,-2,\ldots, -L$) are occupied:
\begin{align}
  \label{eq:95}
    \ket{\Omega} = \prod_{j=-L}^{-1} b_{j}\dg\ket{\emptyset}. 
\end{align}

By definition, $\ket{\Omega}$ is a plugged state. Other plugged states consist of particle-hole excitations from the ground state that leave the bound level occupied. The number $k$ of particle-hole excitations ranges from $k=1$ to $k=L-1$.

With this in mind, we can associate each  plugged state $\ket{\rp}$ containing $k$ particle-hole excitations ($k=1,\ldots,L-1$) with two unique sets of $k$ integers:
\begin{align}
  \label{eq:96}
  \setp &\equiv \{p_1\z, p_2\z,\ldots,p_k\z\} \nonumber \\ (0\le p_1\z< &p_2\z<\ldots< p_k\le L-1),
\end{align}
and
\begin{align}\label{eq:97}
\seth &\equiv \{h_1\z, h_2\z,\ldots,h_k\z\} \nonumber \\ (-1\ge h_1\z &> h_2\z>\ldots>h_k\z> -L),
\end{align}
so that the energy levels indexed by the $p_j\z$ ($j=1, \ldots, k$) lie above the Fermi level and are filled, while the levels indexed by $h_j\z$ lie below and are vacant. 

With $h_k\z > -L$, the bound level is occupied. The labels in Eq.~\eqref{eq:96}~and \eqref{eq:97} are ordered so that $p_1\z$ and $h_1\z$ are closest in energy. If we exclude these two labels,  $p_2\z$ and $h_2\z$ will be closest in energy, and so forth until $p_k\z$ and $h_k\z$. Alternatively, the two equations establish a sequence of electron-hole pairings in increasing energy order.

Given the sets $\setp$ and $\seth$, the eigenstate $\ket{\rp}$ can be written as
\begin{align}
  \label{eq:98}
  \ket{\rp} = \prod_{j=1}^k b_{p_j}\dg b_{h_j}\ket{\Omega},
\end{align}
and its energy, measured from the ground state, as
\begin{align}
  \label{eq:100}
  E_{r_P\z}\z = \dl\sum_{j=1}^k (p_j\z-h_j\z).
\end{align}
where we used Eq.~\eqref{eq:91_}. 

It will be helpful to define a third set $\{g\}$. To this end, let $\{g_0\z\}\equiv\{-1, -2, \ldots, -L\}$ and define $\{g\}$ as the complement of $\{h\}$ with respect to $\{g_0\z\}$. In other words, whereas $\{h\}$ contains the $k$ labels of the negative energy levels that are vacant in $\ket{\rp}$,  the set $\{g\}$ contains the labels of the $L-k$ negative energy levels that remain occupied. Like the levels in $\{h\}$, the elements of $\{g\}$ are arranged in descending energy order. Since $\ket{r_P\z}$ is a plugged state, $g_{L-k}\z= -L$.

From Eq.~\eqref{eq:98}, it follows that the projection of $\ket{\rp}$ onto the initial ground state is
  \begin{align}
    \label{eq:101}
    \braket{r_P\z}{\Psi_0\z} =
    \mel{\Omega}{\prod_{j=1}^kb_{h_j\z}\dg b_{p_j\z}}{\Psi_0\z}.
  \end{align}

It is instructive first to consider the special case $k=0$, that is, the projection $\langle\Omega|\Psi_0\rangle$. According to Wick's Theorem, this projection is the determinant of an $L\times L$ matrix whose elements are anticommutators between initial- and final-state operators, $a_q\z$ and $b_\ell\z$ ($q,\ell=-L,\ldots,-1$). Among the $L!$ products resulting from the expansion of the determinant, the diagonal term is dominant, since the anticommutators~\eqref{eq:92_} of operators $a_q\z$ and $b_\ell\z$ with $q=\ell$ have a larger absolute value than those of operators with $q\ne\ell$. With $q=\ell$, Eq.~\eqref{eq:92_} reduces to
 \begin{align}\label{eq:102}
 \acomm{b_q\z}{a_q\dg} = \dfrac{\sin(\delta)}{\delta}\qquad(q=-L,\ldots,-1).
 \end{align}
 For $\delta\to0$, the right-hand side of Eq.~\eqref{eq:102} becomes unitary, while all off-diagonal ($q\ne\ell$) anticommutators vanish.
 
 With $k>0$, Wick's Theorem generates determinants with smaller absolute values. The single-particle states $b_{p_j\z}\z$ ($j=1,\ldots, k$) now have positive energies and cannot pair with initial states having the same indices. The term with the highest absolute value in the expansion of $\braket{\rp}{\Psi_0}$ contains only $L-k$ anticommutators between operators with the same index. It follows that the absolute value $\abs{\braket{\rp}{\Psi_0\z}}$ is smaller than $\abs{\braket{\Omega}{\Psi_0}}$ and that excited states with large $k$ make relatively small contributions to the fidelity. For this reason, our analysis focuses on the final many-body states with small $k$.

For $k>0$, we define the partial many-body projection 
\begin{align}\label{eq:104}
  \braket{\Omega}{\Psi_0\z}_{k}\z \equiv
  \mel{\emptyset}{\prod_{j=1}^{L-k} b_{g_j\z}\z { a_{g_j\z}\dg} } {\emptyset},
\end{align}
which pairs each negative energy level occupied in $\ket{r_P\z}$ with the initial level with the same index.

In addition to these levels, the right-hand side of Eq.~\eqref{eq:101} involves the positive energy final eigenstates $b_{p_j\z}\z$ and the negative energy initial eigenstates $a_{h_j\z}$ ($j=1,2,\ldots, k$). As Eq.~\eqref{eq:92_} shows, the magnitude of the anticommutator between an initial and a final eigenoperator increases in inverse proportion to the associated energy difference. Therefore, to obtain a good approximation to the right-hand side of Eq.~\eqref{eq:101}, we pair the set of $L-k$ occupied negative-energy single-particle levels in the final many-body state $\ket{r_P\z}$ with the set of initial-state single-particle levels with the same indices and pair the remaining $k$ (positive-energy) single-particle levels in the final state with the remaining $k$ levels in the initial state. This association allows contractions between the levels occupied in $\ket{\Psi_0\z}$ and the negative-energy levels closest to them in the initial state. Explicitly, this approximation reads
  \begin{align}
\label{eq:103}
    \braket{\rp}{\Psi_0\z} = \braket{\Omega}{\Psi_0\z}_{k} \bra{\emptyset} \prod_{j=1}^k{b_{p_j\z}}{a_{h_j\z}\dg} \ket{\emptyset},
\end{align}
and the substitution of Eq.~\eqref{eq:92_} for the resulting anticommutators on the right-hand side yields the following equality:
\begin{align}\label{eq:107}
\braket{\rp}{\Psi_0\z} = \braket{\Omega}{\Psi_0\z}_{k}\z &\left(\dfrac{\sin(\delta)}{\pi} \right)^k\det(\underline{\underline{M}}),
\end{align}
%\braket{\rp}{\Psi_0\z} = \braket{\Omega}{\Psi_0\z}_{k}\z &\left(\dfrac{\sin(\delta)}{\pi} \right)^k \times \nonumber \\ &\mathrm{det}\left( \left\{ \frac1{h_{l}\z - p_{j}\z +\frac{\delta}{\pi}} \right\}_k\right)
where $\underline{\underline{M}}$ denotes the $k\times k$ matrix with elements 
\begin{align}
\label{eq:1071}
M_{\ell j}\equiv
\frac1{h_{\ell}\z - p_{j}\z +\frac{\delta}{\pi}}
\quad(1\le \ell,j \le k).
\end{align}

With the help of the Cauchy determinant formula \cite{cauchy1841exercices}, we can obtain an expression analogous to Eq.~\eqref{eq:107} for the projection of the final ground state upon the initial one:
\begin{align}\label{eq:106}
\braket{\Omega}{\Psi_0\z} = \braket{\Omega}{\Psi_0\z}_{k}\z\left(\dfrac{\sin(\delta)}{\delta}\right)^k,
\end{align}
for $k \ll L$.

We now divide Eq.~\eqref{eq:107} by Eq.~\eqref{eq:106} to eliminate the common factor $\braket{\Omega}{\Psi_0\z}_{k}\z (\sin(\delta))^k$:
\begin{align}
\label{eq:108}
     \braket{\rp}{\Psi_0\z} = \braket{\Omega}{\Psi_0\z}\Big(\dfrac{\delta}{\pi}\Big)^k\mathrm{det}(\underline{\underline{M}}).
  \end{align}

To obtain the last factor on the right-hand side of Eq.~\eqref{eq:108}, we rely on the Leibniz formula for the determinants. Once squared, the resulting expression reads
\begin{align}\label{aux_proj^2}
{\det(\underline{\underline{M}})}^2&= \sum_{\sigma} \prod_{j=1}^k \frac{1}{M_{\sigma(j)j\z}^2}\nonumber\\ 
&+ \sum_{\sigma,\sigma' \ne \sigma}\mathrm{sgn}(\sigma)\mathrm{sgn}(\sigma')\prod_{j=1}^k \frac{1}{M_{\sigma(j)j}\z}\,\,
\frac{1}{M_{\sigma'(j)j}\z },
\end{align}
where $\sigma$ and $\sigma'$ denote permutations of the sequence $1,2,\ldots, k$.

Whereas the summand of the first term on the right-hand side of Eq.~\eqref{aux_proj^2} is always positive, the sign of the summand in the second term depends on the permutations $\sigma$ and $\sigma'$. Therefore, one might expect the magnitude of the second term to be smaller than the first term. Numerical calculations showed that the second term can be dropped from the equality, as its contribution to the fidelity is small.

We can now substitute the right-hand sides of Eq.~(\ref{eq:108}) for $\braket{\rp}{\Psi_0\z}$ and Eq.~\eqref{eq:100} for $E_{r_P\z}\z$ in the first term on the right-hand side of Eq.~(\ref{eq:93}), to determine the contribution~\eqref{eq:45} of the plugged states to the fidelity:
  \begin{align}
    \label{eq:109}
    %    \pp_P(t) = 
    %{\braket{\Omega}{\Psi_0\z}}^2
    %\sum_{r_P\z}\z \left(\dfrac{\delta}{\pi}\right)^{2k} 
    %\prod_{j=1}^k \dfrac{\exp(-i({p_j\z-h_j\z})\oml t)}{\left(h_{j}\z-p_{j}\z+\dfrac{\delta}{\pi}\right)^2} ,
    \pp_P(t) = 
    {\braket{\Omega}{\Psi_0\z}}^2
    \sum_{r_P\z} &\sum_{\sigma}\z \left(\dfrac{\delta}{\pi}\right)^{2k} 
    \nonumber \\ & \prod_{j=1}^k  \dfrac{\exp(-i({p_j\z-h_j\z})\oml t)}{\left(h_{\sigma(j)}\z-p_{j}\z+\dfrac{\delta}{\pi}\right)^2} ,
  \end{align}
  where
  \begin{align}
    \label{eq:110}
    \oml \equiv \dfrac{\dl}{\hbar} = \dfrac{2\pi}{T_L}.
  \end{align}

Each many-body eigenstate $\ket{\rp}$ is characterized by a number $k$ of particle-hole excitations and the sets $\setp$ and $\setk$ in Eqs.~\eqref{eq:96}~and \eqref{eq:97}, respectively. Therefore, summing over $\ket{\rp}$ amounts to summing over $k$ and, for each $k$, summing over all distributions of $p_j\z$ and $h_j\z$ consistent with Eqs.~\eqref{eq:96}~and \eqref{eq:97}, respectively. 
 
The set $\setp$ can be regarded as a vector $\mathbf{p}$ with components $p_1\z, p_2\z,\ldots, p_k\z$ in a $k$-dimensional vector space $\mathcal{V}_P\z$. The inequalities on the right-hand side of Eq.~(\ref{eq:96}) restrict $\mathbf{p}$ to a subsector of $\mathcal{V}_P\z$. The strict inequalities also exclude the boundaries of the subspace; however, this restriction is inconsequential, as the number of points in the boundaries is $\order{1/L}$, relative to the number of points in $\mathcal{V}_P\z$.

  We can freely interchange the labeling of any pair $(p_i\z,\ p_j\z)$ of elements in $\setp$, provided that we also interchange the labeling of the corresponding pair $(h_i\z,\  h_j\z)$ in $\seth$. Therefore, we can enumerate all permutations of the elements in $\setp$ to form a class of $k!$ vectors that span the entire space $\mathcal{V}_p\z$. This enumeration allows us to eliminate the inequalities so that Eqs.~(\ref{eq:96})~and (\ref{eq:97}) become
  \begin{align}
    \label{eq:111}
    \setp \equiv \{p_1\z, p_2\z,\ldots,p_k\z\}\qquad
    (1\le p_j\z\le L;\ j=1,\ldots, L), 
  \end{align}
  and
\begin{align}
  \seth \equiv \{h_1\z, h_2\z,\ldots,h_k\z\}\qquad
    (-L< h_j\z\le -1;\ j=1,\ldots, L), 
  \end{align}
  respectively.
  
These simplifications bring Eq.~(\ref{eq:109}) to the form
    \begin{align}
    \pp_P(t) = 
    {\braket{\Omega}{\Psi_0\z}}^2 &\sum_{k=0}^L \dfrac{1}{k!}\left(\dfrac{\delta}{\pi}\right)^{2k}\prod_{j=1}^k \nonumber \\ &\sum_{p_j\z=0}^{L-1}\sum_{h_j\z=-(L-1)}^{-1} \frac{\exp(-i(p_j\z-h_j\z)\oml t )}{\left(h_{j}\z-p_{j}\z+\frac{\delta}{\pi}\right)^2}. 
      \label{eq:112}
    \end{align}

    For large $L$, the Euler-Maclaurin summation formula converts the two sums on the right-hand side of Eq.~(\ref{eq:112}) into integrals. This approximation yields the expression
    \begin{align}
    \label{eq:113}
    \pp_P(t) = {\braket{\Omega}{\Psi_0\z}}^2&\sum_{k=0}^L\dfrac{1}{k!}\left(\dfrac{\delta}{\pi}\right)^{2k}\prod_{j=1}^k   \nonumber \\ 
    &\int_{0}^{L-1}\int_{-L+1}^{-1} \frac{\exp(-i(p-h)\oml t)}{\left(h\z-p\z+\frac{\delta}{\pi}\right)^2}\,\dd h\,\dd p,
    \end{align}
or with the substitution $q\equiv h-p+\sfrac{\delta}{\pi}$ for the inner integration variable,
    \begin{align}\label{eq:114}
    \pp_P(t) = {\braket{\Omega}{\Psi_0\z}}^2 &\sum_{k=0}^L \dfrac{1}{k!}\left(\frac{\delta}{\pi}\right)^{2k}\prod_{j=1}^k e^{i\oml \frac{\delta}{\pi} t} \times \nonumber \\
    &\int_{0}^{L-1}\int_{-L+1-p+\frac{\delta}{\pi}}^{-1-p+\frac{\delta}{\pi}}\dfrac{e^{i  q\oml t}}{q^2}\,\dd q\,\dd p. 
    \end{align}

    To compute the integral over $q$ on the right-hand side of Eq.~(\ref{eq:114}), it is expedient to define the function
    \begin{align}
      \label{eq:115}
      \Phi(q)\equiv-\dfrac{e^{iq\oml t}}{q},
    \end{align}
    and differentiate it with respect to $q$:
    \begin{align}\label{eq:116}
      \dv{\Phi}{q} = \dfrac{e^{i q\oml t}}{q^2} - i\oml t \dfrac{e^{iq\oml t}}{q}.
    \end{align}

   From Eq.~(\ref{eq:110}), we have that $t\ll T_L\z$ implies $\oml t\ll1$, which justifies the withdrawal of the second term on the right-hand side of Eq.~\eqref{eq:116}.  This approximation turns $\Phi(q)$ into a primitive of the integrand on the right-hand side of Eq.~(\ref{eq:114}). The inequality $\oml t\ll1$ also makes the exponential multiplying the integrals on the right-hand side approximately unitary and leads to
    \begin{align}
      \label{eq:117}
                \pp_P(t) =& 
    {\braket{\Omega}{\Psi_0\z}}^2
    \sum_{k=0}^L \dfrac{1}{k!}\left(\frac{\delta}{\pi}\right)^{2k} 
    \prod_{j=1}^k  \int_{0}^{L-1} \Bigg( \nonumber \\ &\Phi\left(-1-p+\frac{\delta}{\pi}\right)  - \Phi\left(-L+1 -p +\frac{\delta}{\pi}\right) \Bigg) \dd p. 
    \end{align}

    Equation~(\ref{eq:115}) shows that the second term within parentheses in the integrand on the right-hand side of Eq.~\eqref{eq:117} vanishes as $L\to\infty$ and can be ignored. Using $r= p+1-\sfrac{\delta}{\pi}$ as the integration variable, we have that
    \begin{align}
      \label{eq:118}
                      \pp_P(t) = 
      {\braket{\Omega}{\Psi_0\z}}^2
      \sum_{k=0}^L \dfrac{1}{k!}\left(\frac{\delta}{\pi}\right)^{2k} 
      \left(\int_{1-\frac{\delta}{\pi}}^{L-\frac{\delta}{\pi}}\dfrac{e^{-ir\oml t}}{r}\dd r\right)^k.
    \end{align}
    
Since the summand on the right-hand side of Eq.~\eqref{eq:118} decays rapidly with $k$, it is safe to extend the upper limit to infinity and convert the sum into an exponential, and this reduces the equality to the form
\begin{align}
\label{eq:1210}
\pp_P(t) = 
 {\braket{\Omega}{\Psi_0\z}}^2
      \exp(\left(\frac{\delta}{\pi}\right)^2 
      \int_{1-\frac{\delta}{\pi}}^{L-\frac{\delta}{\pi}}\dfrac{e^{-ir\omega_L\z t}}{r}\dd r).
    \end{align}

At $t=0$, Eq.~\eqref{eq:1210} reads
\begin{align}\label{eq:1211}
\fp(0) = {\braket{\Omega}{\Psi_0\z}}^2
      \exp(\left(\frac{\delta}{\pi}\right)^2 
      \int_{1-\frac{\delta}{\pi}}^{L-\frac{\delta}{\pi}}\dfrac{1}{r}\dd r),
\end{align}
which implies that
\begin{align}\label{eq:1212}
\fp(t) = \fp(0) \exp(-\left(\frac{\delta}{\pi}\right)^2 
      \int_{1-\frac{\delta}{\pi}}^{L-\frac{\delta}{\pi}}\dfrac{1-e^{-ir\omega_L\z t}}{r}\dd r).
\end{align}

Since $\sfrac{\delta}{\pi}\ll L$, we can substitute $L$ for the upper limit of the integral on the right-hand side. We then change the integration variable to $x=r\omega_L\z t$ and recall that $L\omega_L\z t=\pi\omtau$ to rewrite Eq.~\eqref{eq:1212} in the equivalent form
\begin{align}
   \label{eq:123}
\pp_P(t) = \pp_P(0)&\exp(-\left(\frac{\delta}{\pi}\right)^2\int_{0}^{\pi\omtau t}\dfrac{1-\cos(x)}{x}\dd x) \times \nonumber  \\
&\exp(-i\left(\frac{\delta}{\pi}\right)^2 \int_{0}^{\pi\omtau t}\dfrac{\sin(x)}{x}\dd x),
 \end{align}
or, with the shorthand~(\ref{eq:57}), which defines $\tf$ as the exponential of the first integral on the right-hand side, 
    \begin{align}\label{eq:124}
      \pp_P(t) = \pp_P(0)
      \tf^{-\left(\frac{\delta}{\pi}\right)^2} 
      \exp(-i\left(\frac{\delta}{\pi}\right)^2 
      \int_{0}^{\pi\omtau t}\dfrac{\sin(x)}{x}\dd x).
    \end{align}

    \subsection{Unplugged eigenstates}
    \label{sec:unpl-eigenst}

The ground state $\ket{\Omega}$ of the final Hamiltonian is a plugged state by definition. In the lowest-energy unplugged state $\ket{\Omega_U\z}$, electrons occupy the first level above the Fermi level and all levels below it, except the bound state:
    \begin{align}
      \label{eq:125}
      \ket{\Omega_U\z} = b_{1}\dg b_{-1}\dg\ldots b_{-L+1}\dg\ket{\emptyset}.
    \end{align}

As explained in Sec.~\ref{sec:numerical-results}, the energy difference between $\ket{\Omega_U\z}$ and $\ket{\Omega}$ is $\epsilon_B$, that is, the difference between the first positive single-particle energy and the bound-state energy. To construct the other unplugged states, we consider all excitations of $\ket{\Omega_U\z}$ containing $k$ particles and $k$ holes ($k=1,2,\ldots,L-1$). Like $\ket{\rp}$, a generic unplugged state $\ket{\ru}$ with $k$ particle-hole excitations is described by two sets of labels: $\setp=\{1 < p_1 < p_2 <\ldots < p_{k}  \le L\}$ and $\seth=\{-L < h_1 < h_2 <\ldots < h_{k}  \le 1.\}$, in analogy with Eqs.~\eqref{eq:96}~and \eqref{eq:97}, respectively. The following equality then expresses the unplugged state: 
\begin{align}
\label{eq:129}
    \ket{\ru} = \prod_{j=1}^kb_{p_j\z}\dg b_{h_j\z}\z\ket{\Omega_U\z}.
\end{align}

Its energy, measured from the ground state, is
    \begin{align}
      \label{eq:128}
      E_{r_U\z}\z = \epsilon_B\z + \dl\sum_{j=1}^k(p_j\z-h_j\z),
    \end{align}
and its projection over the initial ground state is
\begin{align}
      \label{eq:130}
\braket{\ru}{\Psi_0\z} = \mel{\Omega_U\z}{\prod_{j=1}^k{b_{h_j\z}\dg b_{p_j\z}}}{\Psi_0\z}.
\end{align}

For $k=0$, the right-hand side of Eq.~\eqref{eq:130} reduces to the overlap $\braket{\Omega_U\z}{\Psi_0\z}$, given by Wick's Theorem, as before. However, we can no longer contract the initial single-particle eigenvectors with the bound level, which is vacant. Thus, the lowest diagonal contraction pairs $a_{-L}\dg$ with $b_{-L+1}\z$; $a_{-L+1}\dg$ with $b_{-L}\z$; and so on until $a_{-1}\dg$ with $b_1\z$. The corresponding anticommutators, given by Eq.~(\ref{eq:92_}), are
    \begin{align}
      \label{eq:131}
      \acomm{b_{q+1}\z}{a_{q}\dg} = \dfrac{\sin(\delta)}{\delta - \pi}\qquad(q=-L,\ldots,-1). 
    \end{align}

Equation~\eqref{eq:128}, \eqref{eq:130}~and \eqref{eq:131} are analogous to Eqs.~\eqref{eq:100}, \eqref{eq:101}~and \eqref{eq:102}, respectively. Keeping in mind the distinction between the former trio and the latter, we can start at Eq.~\eqref{eq:130} and follow the algebraic sequence from Eq.~\eqref{eq:101}~to \eqref{eq:124} to derive the following expression for the contribution~\eqref{eq:48} of the unplugged states to the fidelity:
\begin{align}\label{eq:132}
  \pp_U(t) = \pp_U(0)&
      \tf^{-\left(1-\Gusfrac{\delta}{\pi}\right)^2} \times \nonumber \\
      &\exp(-i\left(1-\frac{\delta}{\pi}\right)^2 
      \int_{0}^{\pi\omtau t}\dfrac{\sin(x)}{x}\dd x). 
\end{align}
%where
%\begin{align}\label{eq:133}
% \pp_U(0) = {\braket{\Omega_U\z}{\Psi_0\z}}^2
%      L^{\mathlarger{\left(1-\frac{\delta}{\pi}\right)^2}}.
%\end{align}

\subsection{Fidelity}
\label{sec:appFidel}

To exploit the similarity between Eqs.~(\ref{eq:123})~and (\ref{eq:135}) for $\pp_P(t)$ and $\pp_U(t)$, it is useful to define the ratio
\begin{align}\label{eq:134}
Q \equiv \dfrac{\pp_U(0)}{\pp_P(0)},
\end{align}
and the function
\begin{align}
  \label{eq:137}
 \vartheta_t\z = \left(1-\dfrac{2\delta}{\pi}\right)\sii(\pi\omtau t),
\end{align}
so that Eq.~\eqref{eq:132} can be written as
\begin{align}\label{eq:135}
\pp_U(t) = Q\tf^{-\mathlarger{\left(1-\frac{2\delta}{\pi}\right)}}\exp(-i\vartheta_t\z)
  \pp_P(t), 
\end{align}
or
\begin{align}
\pp_U(t) = \dfrac{\rt}{1+Q} e^\mathlarger{-i \vartheta_t\z} ~ {\tf}^{-\mathlarger{\frac{\alpha}{2}}} e^\mathlarger{-i\frac{\alpha}2\sii(\pi\omtau t)}.
\end{align}

We can then substitute the right-hand sides of Eqs.~(\ref{eq:123})~and (\ref{eq:135}) in Eq.~(\ref{eq:45}) to obtain the fidelity
\begin{align}
  \label{eq:138}
\ff(t) = \abs{1+\rt \exp(-i(\omb t +\vartheta(t) ))}^2\tf^{-2(\frac{\delta}{\pi})^2}\abs{\pp_P(0)}^2,
\end{align}
where $\rt$, defined by Eq.~\eqref{eq:58}, is the ratio between $Q\tf^{-\beta}$ and $\tf^{-\alpha}$.

From the definitions of $\ff(t)$ in Eq. \eqref{eq:44}, $\tf$ in Eq. \eqref{eq:55}, $\rt$ in Eq. \eqref{eq:56}, and $\vartheta_{t}\z$ in Eq. \eqref{eq:57}, we have $\ff(0)= \mathcal{T}_{0}\z=1$, $\mathcal{R}_{0}\z= Q$, and $\vartheta_{0}\z=0$. With these initial values, Eq.~(\ref{eq:138}) implies that
\begin{align}
  \label{eq:139}
  \abs{\pp_P(0)}^2 = \dfrac{1}{(1+ Q)^2},
\end{align}
and substitution of the right-hand side for the last factor in Eq.~\eqref{eq:138} yields:
\begin{align}
  \ff(t) = \dfrac{1 + \mathcal{R}_t^2 + 2\rt\cos(\omb t+\vartheta_t\z)}{(1+ Q)^2} \mathcal{T}_t^{-\alpha}.
\end{align}

\subsection{The ratio $Q$}
\label{sec:q}
Equation~\eqref{eq:1211} relates the initial value $\fp({0})$ of the plugged-state fidelity to the ground state projections ${\braket{\Omega}{\Psi_0\z}}^2$. When we substitute $L$ for $L-\sfrac{\delta}{\pi}$ in the upper limit and evaluate the integral on the right-hand side, it results that
\begin{align}\label{eq:140}
\fp(0) = {\braket{\Omega}{\Psi_0\z}}^2
      \left(\dfrac{L}{1-\frac{\delta}{\pi}}\right)^{\mathlarger{\left(\frac{\delta}{\pi}\right)^2}}.
\end{align}

The analogous expression for the initial value $\fu(0)$ of the unplugged state fidelity is
\begin{align}\label{eq:141}
\fu(0) = {\braket{\Omega_U\z}{\Psi_0\z}}^2
      \left(\dfrac{L}{2-\frac{\delta}{\pi}}\right)^{\mathlarger{\left(1-\frac{\delta}{\pi}\right)^2}}.
\end{align}

From Eq.~\eqref{eq:134} it then follows that
\begin{align}\label{eq:142}
Q = \dfrac{\left(1-\frac{\delta}{\pi}\right)^{\mathlarger{\left(\frac{\delta}{\pi}\right)^2}}}
{\left(2-\frac{\delta}{\pi}\right)^{\mathlarger{\left(1-\frac{\delta}{\pi}\right)^2}}}
\dfrac{{\braket{\Omega_U\z}{\Psi_0\z}}^2}{{\braket{\Omega}{\Psi_0\z}}^2} {L}^{\left(1-2\Gusfrac{\delta}{\pi}\right)},
\end{align}
or more briefly
\begin{align}\label{eq:143}
Q = \dfrac{\left(1-\frac{\delta}{\pi}\right)^{\mathlarger{\frac{\alpha}2}}}
{\left(2-\frac{\delta}{\pi}\right)^{\mathlarger{\frac{\beta}2}}}
\dfrac{{\braket{\Omega_U\z}{\Psi_0\z}}^2}{{\braket{\Omega}{\Psi_0\z}}^2}
{L}^{\mathlarger{\frac{\beta-\alpha}2}}.
\end{align}

The first factor on the right-hand side depends only on the phase shift. The second term depends on $\delta$ and $L$: since the latter is the number of conduction electrons, the denominator of the fraction can be written $AL^{-\sfrac{\alpha}2}$, where the prefactor $A$ depends only on $\delta$ \cite{1967And1049}. The substitution $\delta\to\delta-\pi$ transforms $\alpha$ into $ \beta$. At the same time, it maps the ground state $\ket{\Omega}$ on the lowest-energy unplugged state $\ket{\Omega_U\z}$. Therefore, the denominator ${\braket{\Omega}{\Psi_0\z}}^2= L^{-\sfrac{\alpha}2}$ is mapped on the numerator ${\braket{\Omega_U\z}{\Psi_0\z}}^2= BL^{-\sfrac{\beta}2}$, where $B$ is also independent of $L$, and so the product of the last two terms is $\sfrac{B}{A}$. The ratio $Q$ depends only on $\delta$.

\subsection{Mapping between the plugged and unplugged contributions to the fidelity}\label{sec:T_P->U}

Equation~\eqref{eq:102} expresses the diagonal projections between the initial single-particle eigenstates and the corresponding single-particle eigenstates in a plugged many-body state. The equivalent expression describing the projections on eigenstates in an unplugged many-body state is Eq.~\eqref{eq:131}. The transformation $\delta\to\delta-\pi$ changes the right-side of the former into that of the latter equation, except for an immaterial change of sign. Given the one-to-one correspondence between plugged and unplugged states demonstrated in Fig.~\eqref{fig:3}, we can see that the same transformation maps the contribution of the plugged states to the fidelity onto the contribution of the unplugged states:
\begin{align}\label{AUx_1}
P \to U: \delta \to \delta - \pi.
\end{align}

Conversely, the transformation $\delta\to \pi-\delta$ converts the right-hand side of Eq.~\eqref{eq:131} into that of Eq.~\eqref{eq:102} and hence maps the contribution of the unplugged states onto that of the plugged states:
\begin{align}\label{AUx_2}
U \to P : \delta \to \pi - \delta.
\end{align}

Altogether, the transformations~\eqref{AUx_1}~and \eqref{AUx_2} swap the contributions $fp(t)$, from the plugged states, and $\fu(t)$, from the unplugged states. It follows that the transformations map the ratio $Q$ onto its reciprocal:
\begin{align}\label{AUx_3}
Q = \frac{\fu(0)}{\fp(0)} \to  \frac{\fp(0)}{\fu(0)} = Q^{-1}.
\end{align}

%As mentioned in the conclusions drawn in Sec.~\ref{sec:numerical-results}, the transformation $\alpha \to \beta$ transform the contributions for the fidelity from the plugged states into the contributions of the unplugged states. To bring this point home, from the result in Eq. \eqref{eq:139}, Eq. \eqref{eq:124} can be written as
%\begin{align}\label{AUx_1}
%\pp_{P,\delta}\z(t) = \dfrac{1}{1+Q}  {\tf}^{-\mathlarger{\frac{\alpha}{2}}}\exp(-i\frac{\alpha}2\sii(\pi\omtau t)), 
%\end{align}
%and Eq. \eqref{eq:135} as
%\begin{align}\label{Aux_2}
%\pp_{U,\pi-\delta}\z(t) =  \dfrac{Q}{1+Q}{\tf}^{-\mathlarger{\frac{\beta-\alpha}{2}}} e^\mathlarger{-i \vartheta_t} %\pp_{P,\delta} (t). 
%\end{align}

%Note now that after the transformation $\alpha\mapsto\beta$, the ratio $Q$ in Eq. \eqref{eq:142} is transformed as $Q\mapsto 1/Q$, and the Equation~\eqref{AUx_1} is therefore transformed to
%\begin{align}\label{Aux_3}
% \pp_{U,(\delta-\pi)}\z(t)\equiv \dfrac{Q}{1+Q} {\tf}^{-\mathlarger{\frac{\beta}{2}}}
% \exp(-i\frac{\beta}2
%\sii(\pi\omtau t)),
%\end{align}
%which with the help of Eqs.~\eqref{eq:56}~and \eqref{eq:57} can be rewritten as Eq. \eqref{Aux_2}. This verifies that the transformation $\alpha \mapsto \beta$ indeed maps the contributions from the plugged states to those of the unplugged states.

%Substitution of $\pp_{P,\delta}\z(t)$ for $\fp$ and of $\pp_{U,(\delta-\pi)}\z(t)$ for $\fu$ on the right-hand side of Eq.~\eqref{eq:44} then yields Eq.~\eqref{eq:54}.

%\end{document}

%%% Local Variables:
%%% mode: latex
%%% TeX-master: t
%%% End:

\bibliography{ref.bib,2024boxReference.bib}
\end{document}